\documentclass[aps,prl,reprint,superscriptaddress,
amsfonts,amssymb,amsmath,a4paper,floats,floatfix]{revtex4-2}
\usepackage{graphicx}
\usepackage{siunitx}
\usepackage[dvipsnames]{xcolor}

\begin{document}


\newcommand{\TDEG}{2DEG}

\newcommand{\Iac}{$I_\mathrm{ac}$}
\newcommand{\muS}{$\mu_\mathrm{S}$}
\newcommand{\muI}{$\mu_\mathrm{I}$}
\newcommand{\muD}{$\mu_\mathrm{D}$}
\newcommand{\Vxy}{$V_\mathrm{xy}$}
\newcommand{\VI}{$V_\mathrm{I}$}
\newcommand{\VD}{$V_\mathrm{D}$}
\newcommand{\Leq}{$L_\mathrm{eq}$}
\newcommand{\nuI}{$\nu_\mathrm{I}$}
\newcommand{\nuD}{$\nu_\mathrm{D}$}
\newcommand{\nuB}{$\nu_\mathrm{B}$}
\newcommand{\nB}{$n_\mathrm{B}$}
\newcommand{\nI}{$n_\mathrm{I}$}
\newcommand{\nD}{$n_\mathrm{D}$}
\newcommand{\VBG}{$V_\mathrm{BG}$}


\title{Spin-selective equilibration among integer quantum Hall edge channels}

\author{Giorgio\ Nicol\'i}
	\email{gnicoli@phys.ethz.ch}
	\affiliation{Solid State Physics Laboratory, ETH Z\"urich, 8093 Z\"urich, Switzerland}

\author{Christoph\ Adam}
	\affiliation{Solid State Physics Laboratory, ETH Z\"urich, 8093 Z\"urich, Switzerland}

\author{Marc P.\ R\"o\"osli}
	\affiliation{Solid State Physics Laboratory, ETH Z\"urich, 8093 Z\"urich, Switzerland}
			
\author{Peter\ M\"arki}
	\affiliation{Solid State Physics Laboratory, ETH Z\"urich, 8093 Z\"urich, Switzerland}

\author{Jan\ Scharnetzky}
	\affiliation{Solid State Physics Laboratory, ETH Z\"urich, 8093 Z\"urich, Switzerland}
	
\author{Christian\ Reichl}
	\affiliation{Solid State Physics Laboratory, ETH Z\"urich, 8093 Z\"urich, Switzerland}
	
\author{Werner\ Wegscheider}
	\affiliation{Solid State Physics Laboratory, ETH Z\"urich, 8093 Z\"urich, Switzerland}

\author{Thomas M.\ Ihn}
	\affiliation{Solid State Physics Laboratory, ETH Z\"urich, 8093 Z\"urich, Switzerland}
	
\author{Klaus\ Ensslin}
	\affiliation{Solid State Physics Laboratory, ETH Z\"urich, 8093 Z\"urich, Switzerland}

\date{\today}


\begin{abstract}
The equilibration between quantum Hall edge modes is known to depend on the disorder potential and the steepness of the edge. Modern samples with higher mobilities and setups with lower electron temperatures call for a further exploration of the topic. We develop a framework to systematically measure and analyze the equilibration of many (up to 8) integer edge modes. Our results show that spin-selective coupling dominates even for non-neighboring channels with parallel spin. Changes in magnetic field and bulk density let us control the equilibration until it is almost completely suppressed and dominated only by individual microscopic scatterers. This method could serve as a guideline to investigate and design improved devices, and to study fractional and other exotic states.
\end{abstract}

\maketitle


Quantum Hall devices remain paradigmatic for research on topological systems~\cite{hasan_colloquium_2010-2}. The Hall regime is accessed with a quantizing magnetic field perpendicular to a two-dimensional electron gas (\TDEG)~\cite{klitzing_new_1980-2, tsui_two-dimensional_1982}. Dissipationless non-equilibrium currents flow in one-dimensional chiral channels along the edge of the system in response to an external voltage~\cite{halperin_quantized_1982, prange_quantum_1990-1, haug_edge-state_1993-1, hwang_experimental_1993-2}, experiencing inter-edge-state scattering in the presence of a background disorder potential~\cite{alphenaar_selective_1990-1, haug_scattering_1989-1, komiyama_inter-edge-state_1992, muller_equilibration_1992-1, takagaki_inter-edge-state_1994, washburn_quantized_1988-1, acremann_individual_1999}. Equilibration phenomena among non-equilibrium edge currents are not yet fully understood despite the rich history of past experiments on semiconducting devices.

Haug and coworkers found length-dependent equilibration in spin-degenerate quantum Hall systems with top gates acting as partially transmitting barriers \cite{haug_quantized_1988-3, haug_scattering_1989-1, haug_edge-state_1993-1}, but did not report about spin-related effects. Later, M\"uller found that in the presence of a background disorder potential, spin-orbit interactions mediate the equilibration between spin-polarized edge modes by allowing charge carriers to flip their spin~\cite{muller_equilibration_1992-1, muller_influence_1992}. The continuous advancements in material technologies thus motivated a revival of equilibration experiments~\cite{deviatov_separately_2008-1, grivnin_nonequilibrated_2014, lin_charge_2019-1, maiti_magnetic-field-dependent_2020}.

Local probe experiments by Weis~\textit{et al.} already showed the complexity of the microscopic recostruction of the edge potential~\cite{wei_edge_1998-2, weitz_low-temperature_2000, huels_long_2004, klaffs_eddy_2004}. Further details on the edge could be revealed assuming that the presence of an incompressible region of a specific filling factor between two channels implies weak equilibration.
 
 Quantum Hall edge state equilibration experiments gradually expanded to the fractional regime too, often finding non-trivial edge recontructions and current distributions~\cite{kouwenhoven_selective_1990-2, chang_transport_1992-1, sabo_edge_2017-3}. Graphene is another mature platform for quantum Hall experiments unraveling the role of valley and spin degrees of freedom in equilibration phenomena~\cite{amet_selective_2014-1, wei_mach-zehnder_2017, zimmermann_tunable_2017, kumar_equilibration_2018}.

In this manuscript, we address the question of inter-edge-mode scattering in state-of-the-art devices using electronic transport experiments. The design that we use is inspired by historically well-known experiments, where edge channels can be reflected and transmitted with barrier gates to obtain well-controlled out-of-equilibrium population of edge modes~\cite{haug_quantized_1988-3, muller_edge_1990-2, faist_interior_1991, muller_influence_1992, woodside_imaging_2001}. We study how the excitation of selected integer edge modes is redistributed as they co-propagate and extract the strength of pair-wise coupling among many channels (up to 8). We find that the spin of the modes determines the equilibration at low enough fields, spin-selectively coupling even distant channels in contrast with many findings from the past~\cite{alphenaar_selective_1990-1, muller_equilibration_1992-1, haug_edge-state_1993-1, takagaki_inter-edge-state_1994}. For larger fields the equilibration is almost completely suppressed and mesoscopic impurities dominate the weak equilibration between spin-split channels.


\begin{figure}[t]
\includegraphics[width=0.95\columnwidth]{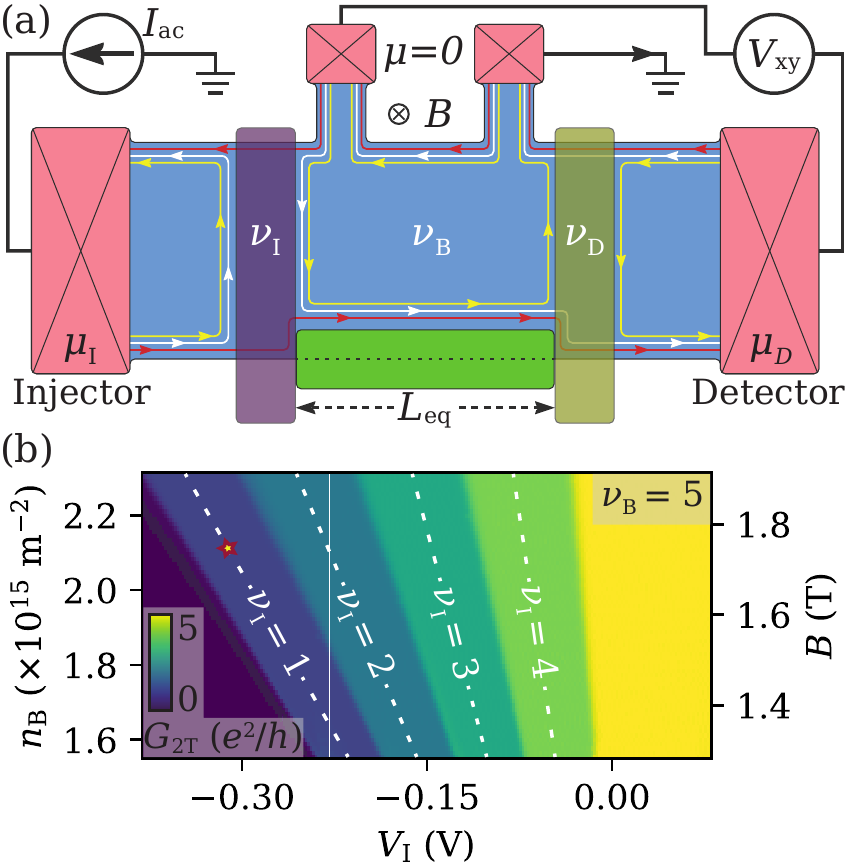}
\caption{\label{fig:sample}(a) Schematic device structure and measurement setup, not to scale. A current $I_\mathrm{ac} = \SI{500}{\pico\ampere}$ flows through the device between injector and ground contacts, at potentials \muI\ and $\mu = 0$ respectively. Two top gates act as barriers downstream of the injector (violet gate) and upstream of the detector (yellow gate) contacts. They are controlled via two dc voltage sources \VI\ and \VD (not shown). A side gate (green) laterally depletes the \TDEG\ along a length $L_\mathrm{eq} = \SI{35}{\micro\meter}$. (b) 2-terminal conductance measured through the device as a function of the left barrier gate voltage \VI . The right barrier gate is grounded~\cite{comm_topgates}. The magnetic field $B$ and bulk density \nB , controlled with the back gate, are stepped together to ensure constant bulk filling factor $\nu_\mathrm{B}=5$. Diagonal dashed lines indicate regions of constant conductance, corresponding to a quantized local filling factor \nuI\ below the injector gate.}
\end{figure}

\begin{figure*}[t]
\includegraphics[width=0.95\textwidth]{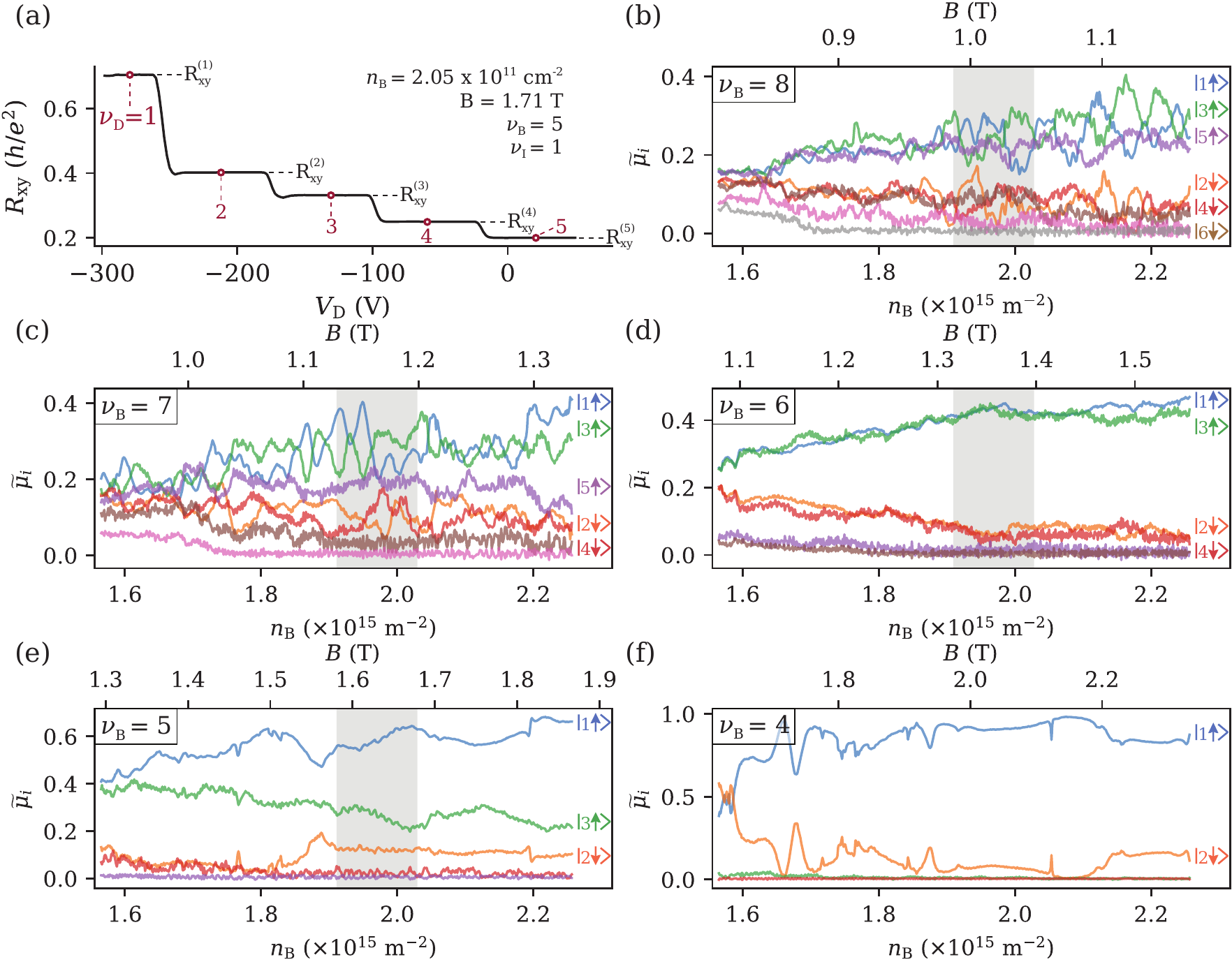}
\caption{\label{fig:mu} (a) Equilibration measurement performed at the star-shaped symbol in Fig~\ref{fig:sample}(b). The red circles indicate the data points required to calculate a set of $\mu_i$ following Eq.~(\ref{eq:rxy}). (b)--(f) Electrochemical potential $\widetilde{\mu}_i$ of the modes for integer bulk filling factors $\nu_\mathrm{B} = \SIrange{8}{4}{}$ after the equilibration path. Matching labels indicate the spin and index of the channels. The back gate voltage \VBG\ (controlling \nB ) and the magnetic field $B$ are simultaneously stepped to fix \nuB\ during each experiment, similar to Fig.~\ref{fig:sample}(b). The shaded regions in figures (a)--(d) indicate the range where coupling parameters have been extracted [see text and Fig.~\ref{fig:coupling}(c)--(f)].}
\end{figure*}

\begin{figure*}[t]
\includegraphics[width=0.95\textwidth]{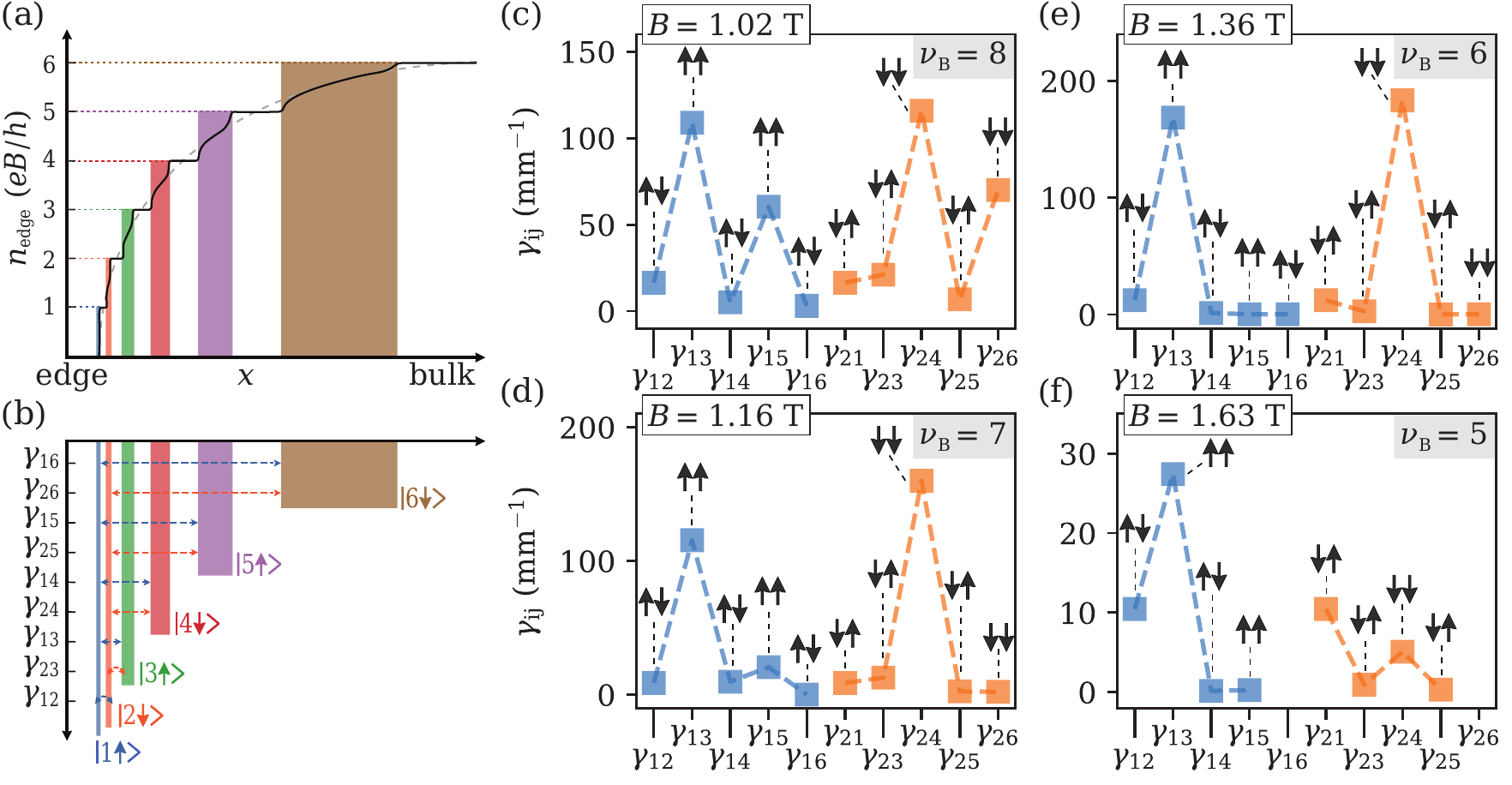}
\caption{\label{fig:coupling}(a) Density profile at the edge of the \TDEG . The unperturbed curve (grey dashed line) was calculated following Ref.~\cite{gerhardts_effect_2008}. The reconstructed density profile is sketched on top with colored shaded regions indicating the compressible stripes. (b) Pictorial representation of channels flowing along one edge of the device. A vertical axis labels the coupling terms represented as horizontal dashed arrows. Even though in our model we consider pair-wise coupling terms $\gamma_{ij}$ between all channels, the sketch shows only terms involving the two outermost channels (1 in blue and 2 in orange). (c)--(f) Coupling parameters for different bulk filling factors. The arrows indicate the spin alignment of the channels coupled by each term. The values reported here are averaged in the grey shaded areas in Fig.~\ref{fig:mu}(b)--(e). The central magnetic field for each shaded area is reported in the figure. The corresponding density range $n_\mathrm{B} = \SIrange{1.91}{2.03}{\times10^{15}\centi\meter^{-2}}$ is the same for each plot.}
\end{figure*}

Our device is an MBE-grown [Al]GaAs heterostructure equipped with a patterned back gate located roughly \SI{1}{\micro\meter} below the plane of the \TDEG~\cite{berl_structured_2016, scharnetzky_novel_2020}. We lithographically defined top gates as indicated in Fig.~\ref{fig:sample}(a). While the back gate tunes the whole device, we locally control the electron density with the injector and detector gates, creating tunable barriers in front of the injector and detector contacts. An additional gate located between the barriers on the side of the device pushes the \TDEG\ away from the physical edge of the mesa and creates a smooth electrostatic edge. A dilution refrigerator lowers the device temperature to $\lesssim \SI{30}{\milli\kelvin}$. From previous characterization, we expect the \TDEG\ to thermalize with the lattice in our setup~\cite{nicoli_quantum_2019, comm_fridge}.

At integer filling factors of the \TDEG, an external magnetic field $B$ induces edge conduction with the chirality indicated in Fig.~\ref{fig:sample}(a). Following the Landauer-B\"uttiker formalism~\cite{buttiker_four-terminal_1986-4, datta_electronic_1995}, the number of channels transmitted by each barrier depends on the local filling factors
\begin{equation}
\label{eq:nu}
\nu_\mathrm{I,D} = \frac{h n_\mathrm{I,D}}{eB} \leq \nu_\mathrm{B},
\end{equation}
where \nuI\ and \nuD\ are the filling factors of the \TDEG\ under the injector and detector gates respectively, when we fix the local densities to \nI\ and \nD . We tune the system such that \nuI , \nuD\ and the bulk filling factor \nuB\ have integer values to perform the experiments in a controlled way and suppress bulk equilibration~\cite{lin_charge_2019-1}. We can route channels carrying different electrochemical potentials to flow along the co-propagation path [\Leq\ in Fig.~\ref{fig:sample}(a)]. Measuring the longitudinal resistance across this path~\cite{haug_edge-state_1993-1, muller_equilibration_1992-1} or the potential of the detector with respect to ground (our case), will yield information about the strength of the equilibration processes among the channels.

We measure the two-terminal conductance $G_\mathrm{2T}$ as a function of the barrier voltage \VI\ while an ac current \Iac\ flows from the injector to ground and the other barrier is fully transparent. Figure~\ref{fig:sample}(b) shows the result measured at $\nu_\mathrm{B} = 5$. Plateaus of constant conductance matching integer multiples of $e^2/h$ are found as the barrier gate voltage decreases, (white dashed lines). Each diagonal feature corresponds to a fixed number of channels transmitted through the barrier region. We repeat the same experiment with the detector gate and for different bulk filling factors to observe the transmission characteristics of both barriers (not shown).

After flowing through the injector barrier, the transmitted channels will have a different electrochemical potential \muI\ compared to the reflected modes on the other side of the barrier, coming from the grounded contact [cf. red versus yellow and white lines in Fig.~\ref{fig:sample}(a)]. However, measuring the transverse voltage \Vxy\ between the detector and ground will reveal no details about inter-mode coupling along the path if all channels equilibrate in the detector ($\nu_\mathrm{D} = \nu_\mathrm{B}$)~\cite{supmat}. The contact settles at the electrochemical potential
\begin{equation}
\label{eq:muD}
\mu_\mathrm{D} = \frac{1}{\nu_\mathrm{D}}\sum_{i=1}^{\nu_\mathrm{D}}\mu_i,
\end{equation}
where $\mu_i$ is the potential of an individual channel $i$ when entering the detector. In the integer regime, all channels have transmission of one and contribute equally to the potential of the contact.  When the detector barrier  allows only selected modes to be transmitted ($\nu_\mathrm{D} < \nu_\mathrm{B}$), measurements of the transverse resistance will yield
\begin{equation}
\label{eq:rxy}
R^{(\nu_\mathrm{D})}_\mathrm{xy} = \frac{V_\mathrm{xy}}{I_\mathrm{ac}} = \frac{h}{e^2}\frac{\mu_\mathrm{D}}{\nu_\mathrm{I}\mu_\mathrm{I}} = \frac{h}{e^2}\frac{1}{\nu_\mathrm{I}\nu_\mathrm{D}}\displaystyle\sum_{i=1}^{\nu_\mathrm{D}}\frac{\mu_i}{\mu_\mathrm{I}}.
\end{equation}

The total equilibration between the channels does not depend on the specific tuning of the barriers, but rather on the edge potential along the propagation path, on the mesoscopic disorder background and on the length of co-propagation. Equilibration among channels under the detector gate does not affect the measurements~\cite{supmat}. If the external current is completely injected in the outermost channel ($\nu_\mathrm{I} = 1$), an out-of-equilibrium population of spin-polarized electrons is built up. This channel can equilibrate either with other channels of the same spin polarization, or with channels of the opposite spin, if spin-flips are involved.

We devised a measurement protocol to extract the electrochemical potential of the channels at the detector. We measure $R^{\nu_\mathrm{D}}_\mathrm{xy}$ for different values of \nuD\ while the injector barrier is tuned to $\nu_\mathrm{I} = 1$. A system of equations of the form of Eq.~(\ref{eq:rxy}) with values $1 \leq \nu_\mathrm{D} \leq \nu_\mathrm{B}$ describes the measurements [see Fig.~\ref{fig:mu}(a)]. We can solve the system to find the normalized electrochemical potentials of the channels $\widetilde{\mu}_i = \mu_i/\mu_\mathrm{I}$, with the initial conditions $\widetilde{\mu}_1^0 = 1$, and $\widetilde{\mu}_j^0 = 0$ for $j \neq 1$

Figures~\ref{fig:mu}(b)--(f) show the results of the analysis for different $\nu_\mathrm{B}$ and in a range of magnetic fields and bulk densities. In Fig.~\ref{fig:mu}(b)--(d), we observe that electrons preferentially equilibrate with states of the same spin, leaving channel 1 to occupy states in modes 3 and 5. Channels labeled with even numbers were mostly decoupled from the only initially excited channel and their potential is closer to the bottom of our energy scale.

In particular, when $\nu_\mathrm{B} = 8$ or $6$, two bundles of channels with opposite spin are resolved and well separated in energy. Even though the clear separation between the two bundles is not visible for the case of $\nu_\mathrm{B} = 7$, also here the system favors spin-selective equilibration. The presence of reproducible fluctuations is likely due to impurities occurring on mesoscopic length scales, modulating the coupling between the modes~\cite{comm_fluctuations}.

If the three spin-up channels in Fig.~\ref{fig:mu}(b) completely equilibrate while the others do not participate, we expect to find $\widetilde{\mu}_{1,3,5} = 1/3 \simeq 0.33$, a case nearly reached at the highest densities. Conversely, if all channels equilibrate, then $\widetilde{\mu}_i = 1/8 = 0.125$ for all of them, which is nearly the case at the lowest densities.

As the number of channels in the bulk decreases with increasing external magnetic field, so does the coupling between them. Figure~\ref{fig:mu}(e) ($\nu_\mathrm{B} = 5$) shows that electrons in channels 1 and 3 are not fully equilibrating along \Leq, contrary to the cases with $\nu_\mathrm{B} > 5$. We observe that the coupling becomes weaker for larger $B$, but spin-selective equilibration still remains the favored process. In Fig.~\ref{fig:mu}(f) ($\nu_\mathrm{B} = 4$) the coupling weakens to the point where $\widetilde{\mu}_1 \approx \widetilde{\mu}_1^0 = 1$ for the whole range. Few mesoscopic features increase the coupling between the two modes in the lowest Landau level, which requires some spin-flip mechanism. Spin-selective equilibration is not observed in this case. Performing the same experiments over a distance $L`\prime_\mathrm{eq} = \SI{535}{\micro\meter}$ reveals full equilibration irrespective of the spin alignment, although the innermost channel remains decoupled~\cite{supmat}.

The density profile at the edge, sketched in the spirit of Ref.~\cite{chklovskii_electrostatics_1992-2} in Fig.~\ref{fig:coupling}(a), guides us in understanding the results of Fig.~\ref{fig:mu}. The edge channels represent discrete conducting regions located where the density has a non-zero gradient. Incompressible stripes with a fixed filling factor separate compressible regions that form at the edge as a result of screening and interactions in the presence of an external $B$ field~\cite{chklovskii_electrostatics_1992-2, lier_self-consistent_1994-1, gerhardts_effect_2008, kovrizhin_equilibration_2011-1}. Decreasing \nuB\ at constant density, by increasing $B$, means that a smaller number of channels spans the density profile, pushing the innermost channels further into the bulk~\cite{chang_transport_1992-1, muller_confinement-potential_1992, wei_edge_1998-2}. Increasing $B$ and \nB\ while keeping \nuB\ constant instead results in wider incompressible stripes and a larger separation between the channels. The magnetic length $\ell_\mathrm{B}$, and consequently the spatial extent of the wavefunction of the edge modes, decreases for stronger fields. Since charge transfer between channels requires wavefunction overlap, a larger distance and stronger confinement can quickly suppress the tunneling probability amplitude.

Electrons can in principle tunnel from one channel to any other, conserving or flipping their spin. The energy transfer between modes can be approximated with a system of rate equations of the form~\cite{supmat}
\begin{equation}
\frac{d\mu_i}{dx} = -\frac{1}{2}\sum_{j \neq i}\gamma_{ij}(\mu_i - \mu_j).
\end{equation}
Here the potentials $\mu_i$ are intended to be position dependent along the equilibration path between injector and detector. The terms $\gamma_{ij} = \gamma_{ji}$ model a uniform coupling between channels $i$ and $j$ [see Fig.~\ref{fig:coupling}(b)]. These parameters encapsulate any equilibration process in our model, giving us a quantity related to the average equilibration lengths $\ell_{ij}^\mathrm{eq} = \gamma_{ij}^{-1}$ between channels $i$ and $j$. We can numerically calculate the whole set of $\gamma_{ij}$ by performing an equal amount of independent measurements at the detector, each time setting the barrier filling factors to integer values such that $\nu_I \leq \nu_D$ and $\nu_I, \nu_D < \nu_B$~\cite{supmat}.
 
Starting with $\nu_\mathrm{B} = 8$ in Fig.~\ref{fig:coupling}(c), we observe that spin-conserving coupling terms dominate, while spin-flip terms can be more than one order of magnitude smaller. Spin-selective tunneling couples not only the spatially closest channels with parallel spin (channels 1 and 3), but also terms like $\gamma_{15}$ and $\gamma_{26}$ are much larger than spin-flip terms coupling nearest-neighbors. This shows that it is more likely for electrons to tunnel a larger distance without flipping their spin rather than tunneling through a thinner barrier undergoing a spin-flip event.

Increasing the field and decreasing \nuB\ at constant density progressively decouples the channels. In Fig.~\ref{fig:coupling}(d) and (e) we observe a reduction of the coupling, in particular of the long-distance terms $\gamma_{15}$ and $\gamma_{26}$. In Fig.~\ref{fig:coupling}(f) the trend continues and also short-range spin-selective coefficients decrease. Finally, for $\nu_\mathrm{B} \leq 4$, all the integer channels are mostly decoupled, either too far removed towards the bulk or limited by the frequency of spin-flip events.


In this manuscript, we analyzed our data based on the well-established edge channel picture of the integer quantum Hall effect, finding that channels with parallel spin selectively couple with each other, while flipping the spin of electrons is much less likely. At low enough fields, spin-conserving tunneling even couples modes separated by several compressible and incompressible stripes instead of only neighboring channels with parallel spin. In general, the equilibration process is influenced by experimental parameters, like magnetic field and temperature~\cite{hirai_dependence_1995-1}, and by sample properties, such as material quality and heterostructure design. Controlling the transfer of particles between channels could lead to the use of edge modes as spin rails to transport well-defined magnetic moments in quantum computation experiments~\cite{roulleau_direct_2008-2, duprez_macroscopic_2019, rosenblatt_energy_2020}. The presence of spin-selective signatures at low field would help to integrate such a technology with others that do not tolerate or require high fields.

In the fractional regime, a precise knowledge of the equilibration length is sought after to improve experiments involving interferometers and other confined systems~\cite{roosli_observation_2020, roosli_fractional_2020}, anyonic statistics~\cite{nakamura_aharonovbohm_2019, bartolomei_fractional_2020}, the thermal conductance of exotic states~\cite{granger_observation_2009-1, venkatachalam_local_2012-1, banerjee_observed_2017-2} and the complex edge reconstruction associated to fractional states like $2/3$~\cite{lafont_counter-propagating_2019, cohen_synthesizing_2019} and $5/2$~\cite{banerjee_observation_2018-2, dutta_novel_2021}. Investigating the fractional quantum Hall regime is a natural next step and we believe that the techniques established in this paper could complement investigations of edge reconstruction and the formation of stripes with fractional filling factor in a variety of materials~\cite{sabo_edge_2017-3}.

\begin{acknowledgments}
We thank M. Shayegan for illuminating discussions. We acknowledge support from the ETH FIRST laboratory and financial support from the National Center of Competence in Research "QSIT---Quantum Science and Technology".
\end{acknowledgments}


\begin{thebibliography}{63}%
\makeatletter
\providecommand \@ifxundefined [1]{%
 \@ifx{#1\undefined}
}%
\providecommand \@ifnum [1]{%
 \ifnum #1\expandafter \@firstoftwo
 \else \expandafter \@secondoftwo
 \fi
}%
\providecommand \@ifx [1]{%
 \ifx #1\expandafter \@firstoftwo
 \else \expandafter \@secondoftwo
 \fi
}%
\providecommand \natexlab [1]{#1}%
\providecommand \enquote  [1]{``#1''}%
\providecommand \bibnamefont  [1]{#1}%
\providecommand \bibfnamefont [1]{#1}%
\providecommand \citenamefont [1]{#1}%
\providecommand \href@noop [0]{\@secondoftwo}%
\providecommand \href [0]{\begingroup \@sanitize@url \@href}%
\providecommand \@href[1]{\@@startlink{#1}\@@href}%
\providecommand \@@href[1]{\endgroup#1\@@endlink}%
\providecommand \@sanitize@url [0]{\catcode `\\12\catcode `\$12\catcode
  `\&12\catcode `\#12\catcode `\^12\catcode `\_12\catcode `\%12\relax}%
\providecommand \@@startlink[1]{}%
\providecommand \@@endlink[0]{}%
\providecommand \url  [0]{\begingroup\@sanitize@url \@url }%
\providecommand \@url [1]{\endgroup\@href {#1}{\urlprefix }}%
\providecommand \urlprefix  [0]{URL }%
\providecommand \Eprint [0]{\href }%
\providecommand \doibase [0]{https://doi.org/}%
\providecommand \selectlanguage [0]{\@gobble}%
\providecommand \bibinfo  [0]{\@secondoftwo}%
\providecommand \bibfield  [0]{\@secondoftwo}%
\providecommand \translation [1]{[#1]}%
\providecommand \BibitemOpen [0]{}%
\providecommand \bibitemStop [0]{}%
\providecommand \bibitemNoStop [0]{.\EOS\space}%
\providecommand \EOS [0]{\spacefactor3000\relax}%
\providecommand \BibitemShut  [1]{\csname bibitem#1\endcsname}%
\let\auto@bib@innerbib\@empty
\bibitem [{\citenamefont {Hasan}\ and\ \citenamefont
  {Kane}(2010)}]{hasan_colloquium_2010-2}%
  \BibitemOpen
  \bibfield  {author} {\bibinfo {author} {\bibfnamefont {M.~Z.}\ \bibnamefont
  {Hasan}}\ and\ \bibinfo {author} {\bibfnamefont {C.~L.}\ \bibnamefont
  {Kane}},\ }\bibfield  {title} {\bibinfo {title} {Colloquium: {{Topological}}
  insulators},\ }\href {https://doi.org/10.1103/RevModPhys.82.3045} {\bibfield
  {journal} {\bibinfo  {journal} {Reviews of Modern Physics}\ }\textbf
  {\bibinfo {volume} {82}},\ \bibinfo {pages} {3045} (\bibinfo {year}
  {2010})}\BibitemShut {NoStop}%
\bibitem [{\citenamefont {v.~Klitzing}\ \emph {et~al.}(1980)\citenamefont
  {v.~Klitzing}, \citenamefont {Dorda},\ and\ \citenamefont
  {Pepper}}]{klitzing_new_1980-2}%
  \BibitemOpen
  \bibfield  {author} {\bibinfo {author} {\bibfnamefont {K.}~\bibnamefont
  {v.~Klitzing}}, \bibinfo {author} {\bibfnamefont {G.}~\bibnamefont {Dorda}},\
  and\ \bibinfo {author} {\bibfnamefont {M.}~\bibnamefont {Pepper}},\
  }\bibfield  {title} {\bibinfo {title} {New {{Method}} for {{High}}-{{Accuracy
  Determination}} of the {{Fine}}-{{Structure Constant Based}} on {{Quantized
  Hall Resistance}}},\ }\href {https://doi.org/10.1103/PhysRevLett.45.494}
  {\bibfield  {journal} {\bibinfo  {journal} {Physical Review Letters}\
  }\textbf {\bibinfo {volume} {45}},\ \bibinfo {pages} {494} (\bibinfo {year}
  {1980})}\BibitemShut {NoStop}%
\bibitem [{\citenamefont {Tsui}\ \emph {et~al.}(y 31)\citenamefont {Tsui},
  \citenamefont {Stormer},\ and\ \citenamefont
  {Gossard}}]{tsui_two-dimensional_1982}%
  \BibitemOpen
  \bibfield  {author} {\bibinfo {author} {\bibfnamefont {D.~C.}\ \bibnamefont
  {Tsui}}, \bibinfo {author} {\bibfnamefont {H.~L.}\ \bibnamefont {Stormer}},\
  and\ \bibinfo {author} {\bibfnamefont {A.~C.}\ \bibnamefont {Gossard}},\
  }\bibfield  {title} {\bibinfo {title} {Two-{{Dimensional Magnetotransport}}
  in the {{Extreme Quantum Limit}}},\ }\href
  {https://doi.org/10.1103/PhysRevLett.48.1559} {\bibfield  {journal} {\bibinfo
   {journal} {Physical Review Letters}\ }\textbf {\bibinfo {volume} {48}},\
  \bibinfo {pages} {1559} (\bibinfo {year} {1982, May 31})}\BibitemShut
  {NoStop}%
\bibitem [{\citenamefont {Halperin}(y 15)}]{halperin_quantized_1982}%
  \BibitemOpen
  \bibfield  {author} {\bibinfo {author} {\bibfnamefont {B.~I.}\ \bibnamefont
  {Halperin}},\ }\bibfield  {title} {\bibinfo {title} {Quantized {{Hall}}
  conductance, current-carrying edge states, and the existence of extended
  states in a two-dimensional disordered potential},\ }\href
  {https://doi.org/10.1103/PhysRevB.25.2185} {\bibfield  {journal} {\bibinfo
  {journal} {Physical Review B}\ }\textbf {\bibinfo {volume} {25}},\ \bibinfo
  {pages} {2185} (\bibinfo {year} {1982, February 15})}\BibitemShut {NoStop}%
\bibitem [{\citenamefont {Prange}\ and\ \citenamefont
  {Girvin}(1990)}]{prange_quantum_1990-1}%
  \BibitemOpen
  \bibfield  {author} {\bibinfo {author} {\bibfnamefont {R.}~\bibnamefont
  {Prange}}\ and\ \bibinfo {author} {\bibfnamefont {S.~M.}\ \bibnamefont
  {Girvin}},\ }\href@noop {} {\emph {\bibinfo {title} {The {{Quantum Hall
  Effect}}}}}\ (\bibinfo  {publisher} {{Springer}},\ \bibinfo {address} {{New
  York}},\ \bibinfo {year} {1990})\BibitemShut {NoStop}%
\bibitem [{\citenamefont {Haug}(1993)}]{haug_edge-state_1993-1}%
  \BibitemOpen
  \bibfield  {author} {\bibinfo {author} {\bibfnamefont {R.~J.}\ \bibnamefont
  {Haug}},\ }\bibfield  {title} {\bibinfo {title} {Edge-state transport and its
  experimental consequences in high magnetic fields},\ }\href
  {https://doi.org/10.1088/0268-1242/8/2/001} {\bibfield  {journal} {\bibinfo
  {journal} {Semiconductor Science and Technology}\ }\textbf {\bibinfo {volume}
  {8}},\ \bibinfo {pages} {131} (\bibinfo {year} {1993})}\BibitemShut {NoStop}%
\bibitem [{\citenamefont {Hwang}\ \emph {et~al.}(1993)\citenamefont {Hwang},
  \citenamefont {Tsui},\ and\ \citenamefont
  {Shayegan}}]{hwang_experimental_1993-2}%
  \BibitemOpen
  \bibfield  {author} {\bibinfo {author} {\bibfnamefont {S.~W.}\ \bibnamefont
  {Hwang}}, \bibinfo {author} {\bibfnamefont {D.~C.}\ \bibnamefont {Tsui}},\
  and\ \bibinfo {author} {\bibfnamefont {M.}~\bibnamefont {Shayegan}},\
  }\bibfield  {title} {\bibinfo {title} {Experimental evidence for finite-width
  edge channels in integer and fractional quantum {{Hall}} effects},\ }\href
  {https://doi.org/10.1103/PhysRevB.48.8161} {\bibfield  {journal} {\bibinfo
  {journal} {Physical Review B}\ }\textbf {\bibinfo {volume} {48}},\ \bibinfo
  {pages} {8161} (\bibinfo {year} {1993})}\BibitemShut {NoStop}%
\bibitem [{\citenamefont {Alphenaar}\ \emph {et~al.}(1990)\citenamefont
  {Alphenaar}, \citenamefont {McEuen}, \citenamefont {Wheeler},\ and\
  \citenamefont {Sacks}}]{alphenaar_selective_1990-1}%
  \BibitemOpen
  \bibfield  {author} {\bibinfo {author} {\bibfnamefont {B.~W.}\ \bibnamefont
  {Alphenaar}}, \bibinfo {author} {\bibfnamefont {P.~L.}\ \bibnamefont
  {McEuen}}, \bibinfo {author} {\bibfnamefont {R.~G.}\ \bibnamefont
  {Wheeler}},\ and\ \bibinfo {author} {\bibfnamefont {R.~N.}\ \bibnamefont
  {Sacks}},\ }\bibfield  {title} {\bibinfo {title} {Selective equilibration
  among the current-carrying states in the quantum {{Hall}} regime},\
  }\href@noop {} {\bibfield  {journal} {\bibinfo  {journal} {Physical review
  letters}\ }\textbf {\bibinfo {volume} {64}},\ \bibinfo {pages} {677}
  (\bibinfo {year} {1990})}\BibitemShut {NoStop}%
\bibitem [{\citenamefont {Haug}\ \emph {et~al.}(1989)\citenamefont {Haug},
  \citenamefont {Kucera}, \citenamefont {Streda},\ and\ \citenamefont {{von
  Klitzing}}}]{haug_scattering_1989-1}%
  \BibitemOpen
  \bibfield  {author} {\bibinfo {author} {\bibfnamefont {R.~J.}\ \bibnamefont
  {Haug}}, \bibinfo {author} {\bibfnamefont {J.}~\bibnamefont {Kucera}},
  \bibinfo {author} {\bibfnamefont {P.}~\bibnamefont {Streda}},\ and\ \bibinfo
  {author} {\bibfnamefont {K.}~\bibnamefont {{von Klitzing}}},\ }\bibfield
  {title} {\bibinfo {title} {Scattering experiments in two-dimensional systems
  in the presence of quantizing magnetic fields},\ }\href@noop {} {\bibfield
  {journal} {\bibinfo  {journal} {Physical Review B}\ }\textbf {\bibinfo
  {volume} {39}},\ \bibinfo {pages} {10892} (\bibinfo {year}
  {1989})}\BibitemShut {NoStop}%
\bibitem [{\citenamefont {Komiyama}\ \emph {et~al.}(1992)\citenamefont
  {Komiyama}, \citenamefont {Hirai}, \citenamefont {Ohsawa}, \citenamefont
  {Matsuda}, \citenamefont {Sasa},\ and\ \citenamefont
  {Fujii}}]{komiyama_inter-edge-state_1992}%
  \BibitemOpen
  \bibfield  {author} {\bibinfo {author} {\bibfnamefont {S.}~\bibnamefont
  {Komiyama}}, \bibinfo {author} {\bibfnamefont {H.}~\bibnamefont {Hirai}},
  \bibinfo {author} {\bibfnamefont {M.}~\bibnamefont {Ohsawa}}, \bibinfo
  {author} {\bibfnamefont {Y.}~\bibnamefont {Matsuda}}, \bibinfo {author}
  {\bibfnamefont {S.}~\bibnamefont {Sasa}},\ and\ \bibinfo {author}
  {\bibfnamefont {T.}~\bibnamefont {Fujii}},\ }\bibfield  {title} {\bibinfo
  {title} {Inter-edge-state scattering and nonlinear effects in a
  two-dimensional electron gas at high magnetic fields},\ }\href@noop {}
  {\bibfield  {journal} {\bibinfo  {journal} {Physical Review B}\ }\textbf
  {\bibinfo {volume} {45}},\ \bibinfo {pages} {11085} (\bibinfo {year}
  {1992})}\BibitemShut {NoStop}%
\bibitem [{\citenamefont {M{\"u}ller}\ \emph
  {et~al.}(1992{\natexlab{a}})\citenamefont {M{\"u}ller}, \citenamefont
  {Weiss}, \citenamefont {Khaetskii}, \citenamefont {{von Klitzing}},
  \citenamefont {Koch}, \citenamefont {Nickel}, \citenamefont {Schlapp},\ and\
  \citenamefont {L{\"o}sch}}]{muller_equilibration_1992-1}%
  \BibitemOpen
  \bibfield  {author} {\bibinfo {author} {\bibfnamefont {G.}~\bibnamefont
  {M{\"u}ller}}, \bibinfo {author} {\bibfnamefont {D.}~\bibnamefont {Weiss}},
  \bibinfo {author} {\bibfnamefont {A.~V.}\ \bibnamefont {Khaetskii}}, \bibinfo
  {author} {\bibfnamefont {K.}~\bibnamefont {{von Klitzing}}}, \bibinfo
  {author} {\bibfnamefont {S.}~\bibnamefont {Koch}}, \bibinfo {author}
  {\bibfnamefont {H.}~\bibnamefont {Nickel}}, \bibinfo {author} {\bibfnamefont
  {W.}~\bibnamefont {Schlapp}},\ and\ \bibinfo {author} {\bibfnamefont
  {R.}~\bibnamefont {L{\"o}sch}},\ }\bibfield  {title} {\bibinfo {title}
  {Equilibration length of electrons in spin-polarized edge channels},\
  }\href@noop {} {\bibfield  {journal} {\bibinfo  {journal} {Physical Review
  B}\ }\textbf {\bibinfo {volume} {45}},\ \bibinfo {pages} {3932} (\bibinfo
  {year} {1992}{\natexlab{a}})}\BibitemShut {NoStop}%
\bibitem [{\citenamefont {Takagaki}\ \emph {et~al.}(1994)\citenamefont
  {Takagaki}, \citenamefont {Friedland}, \citenamefont {Herfort}, \citenamefont
  {Kostial},\ and\ \citenamefont {Ploog}}]{takagaki_inter-edge-state_1994}%
  \BibitemOpen
  \bibfield  {author} {\bibinfo {author} {\bibfnamefont {Y.}~\bibnamefont
  {Takagaki}}, \bibinfo {author} {\bibfnamefont {K.~J.}\ \bibnamefont
  {Friedland}}, \bibinfo {author} {\bibfnamefont {J.}~\bibnamefont {Herfort}},
  \bibinfo {author} {\bibfnamefont {H.}~\bibnamefont {Kostial}},\ and\ \bibinfo
  {author} {\bibfnamefont {K.}~\bibnamefont {Ploog}},\ }\bibfield  {title}
  {\bibinfo {title} {Inter-edge-state scattering in the spin-polarized quantum
  {{Hall}} regime with current injection into inner states},\ }\href@noop {}
  {\bibfield  {journal} {\bibinfo  {journal} {Physical Review B}\ }\textbf
  {\bibinfo {volume} {50}},\ \bibinfo {pages} {4456} (\bibinfo {year}
  {1994})}\BibitemShut {NoStop}%
\bibitem [{\citenamefont {Washburn}\ \emph {et~al.}(1988)\citenamefont
  {Washburn}, \citenamefont {Fowler}, \citenamefont {Schmid},\ and\
  \citenamefont {Kern}}]{washburn_quantized_1988-1}%
  \BibitemOpen
  \bibfield  {author} {\bibinfo {author} {\bibfnamefont {S.}~\bibnamefont
  {Washburn}}, \bibinfo {author} {\bibfnamefont {A.~B.}\ \bibnamefont
  {Fowler}}, \bibinfo {author} {\bibfnamefont {H.}~\bibnamefont {Schmid}},\
  and\ \bibinfo {author} {\bibfnamefont {D.}~\bibnamefont {Kern}},\ }\bibfield
  {title} {\bibinfo {title} {Quantized {{Hall Effect}} in the {{Presence}} of
  {{Backscattering}}},\ }\href {https://doi.org/10.1103/PhysRevLett.61.2801}
  {\bibfield  {journal} {\bibinfo  {journal} {Physical Review Letters}\
  }\textbf {\bibinfo {volume} {61}},\ \bibinfo {pages} {2801} (\bibinfo {year}
  {1988})}\BibitemShut {NoStop}%
\bibitem [{\citenamefont {Acremann}\ \emph {et~al.}(1999)\citenamefont
  {Acremann}, \citenamefont {Heinzel}, \citenamefont {Ensslin}, \citenamefont
  {Gini}, \citenamefont {Melchior},\ and\ \citenamefont
  {Holland}}]{acremann_individual_1999}%
  \BibitemOpen
  \bibfield  {author} {\bibinfo {author} {\bibfnamefont {Y.}~\bibnamefont
  {Acremann}}, \bibinfo {author} {\bibfnamefont {T.}~\bibnamefont {Heinzel}},
  \bibinfo {author} {\bibfnamefont {K.}~\bibnamefont {Ensslin}}, \bibinfo
  {author} {\bibfnamefont {E.}~\bibnamefont {Gini}}, \bibinfo {author}
  {\bibfnamefont {H.}~\bibnamefont {Melchior}},\ and\ \bibinfo {author}
  {\bibfnamefont {M.}~\bibnamefont {Holland}},\ }\bibfield  {title} {\bibinfo
  {title} {Individual scatterers as microscopic origin of equilibration between
  spin-polarized edge channels in the quantum {{Hall}} regime},\ }\href
  {https://doi.org/10.1103/PhysRevB.59.2116} {\bibfield  {journal} {\bibinfo
  {journal} {Physical Review B}\ }\textbf {\bibinfo {volume} {59}},\ \bibinfo
  {pages} {2116} (\bibinfo {year} {1999})}\BibitemShut {NoStop}%
\bibitem [{\citenamefont {Haug}\ \emph {et~al.}(1988)\citenamefont {Haug},
  \citenamefont {MacDonald}, \citenamefont {Streda},\ and\ \citenamefont {{von
  Klitzing}}}]{haug_quantized_1988-3}%
  \BibitemOpen
  \bibfield  {author} {\bibinfo {author} {\bibfnamefont {R.~J.}\ \bibnamefont
  {Haug}}, \bibinfo {author} {\bibfnamefont {A.~H.}\ \bibnamefont {MacDonald}},
  \bibinfo {author} {\bibfnamefont {P.}~\bibnamefont {Streda}},\ and\ \bibinfo
  {author} {\bibfnamefont {K.}~\bibnamefont {{von Klitzing}}},\ }\bibfield
  {title} {\bibinfo {title} {Quantized {{Multichannel Magnetotransport}}
  through a {{Barrier}} in {{Two Dimensions}}},\ }\href
  {https://doi.org/10.1103/PhysRevLett.61.2797} {\bibfield  {journal} {\bibinfo
   {journal} {Physical Review Letters}\ }\textbf {\bibinfo {volume} {61}},\
  \bibinfo {pages} {2797} (\bibinfo {year} {1988})}\BibitemShut {NoStop}%
\bibitem [{\citenamefont {M{\"u}ller}\ \emph
  {et~al.}(1992{\natexlab{b}})\citenamefont {M{\"u}ller}, \citenamefont
  {Diessel}, \citenamefont {Weiss}, \citenamefont {{von Klitzing}},
  \citenamefont {Ploog}, \citenamefont {Nickel}, \citenamefont {Schlapp},\ and\
  \citenamefont {L{\"o}sch}}]{muller_influence_1992}%
  \BibitemOpen
  \bibfield  {author} {\bibinfo {author} {\bibfnamefont {G.}~\bibnamefont
  {M{\"u}ller}}, \bibinfo {author} {\bibfnamefont {E.}~\bibnamefont {Diessel}},
  \bibinfo {author} {\bibfnamefont {D.}~\bibnamefont {Weiss}}, \bibinfo
  {author} {\bibfnamefont {K.}~\bibnamefont {{von Klitzing}}}, \bibinfo
  {author} {\bibfnamefont {K.}~\bibnamefont {Ploog}}, \bibinfo {author}
  {\bibfnamefont {H.}~\bibnamefont {Nickel}}, \bibinfo {author} {\bibfnamefont
  {W.}~\bibnamefont {Schlapp}},\ and\ \bibinfo {author} {\bibfnamefont
  {R.}~\bibnamefont {L{\"o}sch}},\ }\bibfield  {title} {\bibinfo {title}
  {Influence of interedge channel scattering on the magneto-transport of
  {{2D}}-systems},\ }\href {https://doi.org/10.1016/0039-6028(92)90352-7}
  {\bibfield  {journal} {\bibinfo  {journal} {Surface Science}\ }\textbf
  {\bibinfo {volume} {263}},\ \bibinfo {pages} {280} (\bibinfo {year}
  {1992}{\natexlab{b}})}\BibitemShut {NoStop}%
\bibitem [{\citenamefont {Deviatov}\ and\ \citenamefont
  {Lorke}(2008)}]{deviatov_separately_2008-1}%
  \BibitemOpen
  \bibfield  {author} {\bibinfo {author} {\bibfnamefont {E.~V.}\ \bibnamefont
  {Deviatov}}\ and\ \bibinfo {author} {\bibfnamefont {A.}~\bibnamefont
  {Lorke}},\ }\bibfield  {title} {\bibinfo {title} {Separately contacted edge
  states at high imbalance in the integer and fractional quantum {{Hall}}
  effect regime},\ }\href {https://doi.org/10.1002/pssb.200743341} {\bibfield
  {journal} {\bibinfo  {journal} {physica status solidi (b)}\ }\textbf
  {\bibinfo {volume} {245}},\ \bibinfo {pages} {366} (\bibinfo {year}
  {2008})}\BibitemShut {NoStop}%
\bibitem [{\citenamefont {Grivnin}\ \emph {et~al.}(2014)\citenamefont
  {Grivnin}, \citenamefont {Inoue}, \citenamefont {Ronen}, \citenamefont
  {Baum}, \citenamefont {Heiblum}, \citenamefont {Umansky},\ and\ \citenamefont
  {Mahalu}}]{grivnin_nonequilibrated_2014}%
  \BibitemOpen
  \bibfield  {author} {\bibinfo {author} {\bibfnamefont {A.}~\bibnamefont
  {Grivnin}}, \bibinfo {author} {\bibfnamefont {H.}~\bibnamefont {Inoue}},
  \bibinfo {author} {\bibfnamefont {Y.}~\bibnamefont {Ronen}}, \bibinfo
  {author} {\bibfnamefont {Y.}~\bibnamefont {Baum}}, \bibinfo {author}
  {\bibfnamefont {M.}~\bibnamefont {Heiblum}}, \bibinfo {author} {\bibfnamefont
  {V.}~\bibnamefont {Umansky}},\ and\ \bibinfo {author} {\bibfnamefont
  {D.}~\bibnamefont {Mahalu}},\ }\bibfield  {title} {\bibinfo {title}
  {Nonequilibrated {{Counterpropagating Edge Modes}} in the {{Fractional
  Quantum Hall Regime}}},\ }\bibfield  {journal} {\bibinfo  {journal} {Physical
  Review Letters}\ }\textbf {\bibinfo {volume} {113}},\ \href
  {https://doi.org/10.1103/PhysRevLett.113.266803}
  {10.1103/PhysRevLett.113.266803} (\bibinfo {year} {2014})\BibitemShut
  {NoStop}%
\bibitem [{\citenamefont {Lin}\ \emph {et~al.}(2019)\citenamefont {Lin},
  \citenamefont {Eguchi}, \citenamefont {Hashisaka}, \citenamefont {Akiho},
  \citenamefont {Muraki},\ and\ \citenamefont {Fujisawa}}]{lin_charge_2019-1}%
  \BibitemOpen
  \bibfield  {author} {\bibinfo {author} {\bibfnamefont {C.}~\bibnamefont
  {Lin}}, \bibinfo {author} {\bibfnamefont {R.}~\bibnamefont {Eguchi}},
  \bibinfo {author} {\bibfnamefont {M.}~\bibnamefont {Hashisaka}}, \bibinfo
  {author} {\bibfnamefont {T.}~\bibnamefont {Akiho}}, \bibinfo {author}
  {\bibfnamefont {K.}~\bibnamefont {Muraki}},\ and\ \bibinfo {author}
  {\bibfnamefont {T.}~\bibnamefont {Fujisawa}},\ }\bibfield  {title} {\bibinfo
  {title} {Charge equilibration in integer and fractional quantum {{Hall}} edge
  channels in a generalized {{Hall}}-bar device},\ }\href
  {https://doi.org/10.1103/PhysRevB.99.195304} {\bibfield  {journal} {\bibinfo
  {journal} {Physical Review B}\ }\textbf {\bibinfo {volume} {99}},\ \bibinfo
  {pages} {195304} (\bibinfo {year} {2019})},\ \Eprint
  {https://arxiv.org/abs/1905.01126} {arXiv:1905.01126} \BibitemShut {NoStop}%
\bibitem [{\citenamefont {Maiti}\ \emph {et~al.}(2020)\citenamefont {Maiti},
  \citenamefont {Agarwal}, \citenamefont {Purkait}, \citenamefont {Sreejith},
  \citenamefont {Das}, \citenamefont {Biasiol}, \citenamefont {Sorba},\ and\
  \citenamefont {Karmakar}}]{maiti_magnetic-field-dependent_2020}%
  \BibitemOpen
  \bibfield  {author} {\bibinfo {author} {\bibfnamefont {T.}~\bibnamefont
  {Maiti}}, \bibinfo {author} {\bibfnamefont {P.}~\bibnamefont {Agarwal}},
  \bibinfo {author} {\bibfnamefont {S.}~\bibnamefont {Purkait}}, \bibinfo
  {author} {\bibfnamefont {G.~J.}\ \bibnamefont {Sreejith}}, \bibinfo {author}
  {\bibfnamefont {S.}~\bibnamefont {Das}}, \bibinfo {author} {\bibfnamefont
  {G.}~\bibnamefont {Biasiol}}, \bibinfo {author} {\bibfnamefont
  {L.}~\bibnamefont {Sorba}},\ and\ \bibinfo {author} {\bibfnamefont
  {B.}~\bibnamefont {Karmakar}},\ }\bibfield  {title} {\bibinfo {title}
  {Magnetic-{{Field}}-{{Dependent Equilibration}} of {{Fractional Quantum Hall
  Edge Modes}}},\ }\href {https://doi.org/10.1103/PhysRevLett.125.076802}
  {\bibfield  {journal} {\bibinfo  {journal} {Physical Review Letters}\
  }\textbf {\bibinfo {volume} {125}},\ \bibinfo {pages} {076802} (\bibinfo
  {year} {2020})}\BibitemShut {NoStop}%
\bibitem [{\citenamefont {Wei}\ \emph {et~al.}(1998)\citenamefont {Wei},
  \citenamefont {Weis}, \citenamefont {v.~Klitzing},\ and\ \citenamefont
  {Eberl}}]{wei_edge_1998-2}%
  \BibitemOpen
  \bibfield  {author} {\bibinfo {author} {\bibfnamefont {Y.~Y.}\ \bibnamefont
  {Wei}}, \bibinfo {author} {\bibfnamefont {J.}~\bibnamefont {Weis}}, \bibinfo
  {author} {\bibfnamefont {K.}~\bibnamefont {v.~Klitzing}},\ and\ \bibinfo
  {author} {\bibfnamefont {K.}~\bibnamefont {Eberl}},\ }\bibfield  {title}
  {\bibinfo {title} {Edge {{Strips}} in the {{Quantum Hall Regime Imaged}} by a
  {{Single}}-{{Electron Transistor}}},\ }\href
  {https://doi.org/10.1103/PhysRevLett.81.1674} {\bibfield  {journal} {\bibinfo
   {journal} {Physical Review Letters}\ }\textbf {\bibinfo {volume} {81}},\
  \bibinfo {pages} {1674} (\bibinfo {year} {1998})}\BibitemShut {NoStop}%
\bibitem [{\citenamefont {Weitz}\ \emph {et~al.}(2000)\citenamefont {Weitz},
  \citenamefont {Ahlswede}, \citenamefont {Weis}, \citenamefont {v~Klitzing},\
  and\ \citenamefont {Eberl}}]{weitz_low-temperature_2000}%
  \BibitemOpen
  \bibfield  {author} {\bibinfo {author} {\bibfnamefont {P.}~\bibnamefont
  {Weitz}}, \bibinfo {author} {\bibfnamefont {E.}~\bibnamefont {Ahlswede}},
  \bibinfo {author} {\bibfnamefont {J.}~\bibnamefont {Weis}}, \bibinfo {author}
  {\bibfnamefont {K.}~\bibnamefont {v~Klitzing}},\ and\ \bibinfo {author}
  {\bibfnamefont {K.}~\bibnamefont {Eberl}},\ }\bibfield  {title} {\bibinfo
  {title} {A low-temperature scanning force microscope for investigating buried
  two-dimensional electron systems under quantum {{Hall}} conditions},\ }\href
  {https://doi.org/10.1016/S0169-4332(99)00550-4} {\bibfield  {journal}
  {\bibinfo  {journal} {Applied Surface Science}\ }\textbf {\bibinfo {volume}
  {157}},\ \bibinfo {pages} {349} (\bibinfo {year} {2000})}\BibitemShut
  {NoStop}%
\bibitem [{\citenamefont {Huels}\ \emph {et~al.}(2004)\citenamefont {Huels},
  \citenamefont {Weis}, \citenamefont {Smet}, \citenamefont {v.~Klitzing},\
  and\ \citenamefont {Wasilewski}}]{huels_long_2004}%
  \BibitemOpen
  \bibfield  {author} {\bibinfo {author} {\bibfnamefont {J.}~\bibnamefont
  {Huels}}, \bibinfo {author} {\bibfnamefont {J.}~\bibnamefont {Weis}},
  \bibinfo {author} {\bibfnamefont {J.}~\bibnamefont {Smet}}, \bibinfo {author}
  {\bibfnamefont {K.}~\bibnamefont {v.~Klitzing}},\ and\ \bibinfo {author}
  {\bibfnamefont {Z.~R.}\ \bibnamefont {Wasilewski}},\ }\bibfield  {title}
  {\bibinfo {title} {Long time relaxation phenomena of a two-dimensional
  electron system within integer quantum {{Hall}} plateau regimes after
  magnetic field sweeps},\ }\href {https://doi.org/10.1103/PhysRevB.69.085319}
  {\bibfield  {journal} {\bibinfo  {journal} {Physical Review B}\ }\textbf
  {\bibinfo {volume} {69}},\ \bibinfo {pages} {085319} (\bibinfo {year}
  {2004})}\BibitemShut {NoStop}%
\bibitem [{\citenamefont {Klaffs}\ \emph {et~al.}(2004)\citenamefont {Klaffs},
  \citenamefont {Krupenin}, \citenamefont {Weis},\ and\ \citenamefont
  {Ahlers}}]{klaffs_eddy_2004}%
  \BibitemOpen
  \bibfield  {author} {\bibinfo {author} {\bibfnamefont {T.}~\bibnamefont
  {Klaffs}}, \bibinfo {author} {\bibfnamefont {V.~A.}\ \bibnamefont
  {Krupenin}}, \bibinfo {author} {\bibfnamefont {J.}~\bibnamefont {Weis}},\
  and\ \bibinfo {author} {\bibfnamefont {F.~J.}\ \bibnamefont {Ahlers}},\
  }\bibfield  {title} {\bibinfo {title} {Eddy currents in the integer quantum
  {{Hall}} regime spatially resolved by multiple single-electron transistor
  electrometers},\ }\href {https://doi.org/10.1016/j.physe.2003.12.112}
  {\bibfield  {journal} {\bibinfo  {journal} {Physica E: Low-dimensional
  Systems and Nanostructures}\ }\bibinfo {series} {15th {{International
  Conference}} on {{Electronic Propreties}} of {{Two}}-{{Dimensional Systems}}
  ({{EP2DS}}-15)},\ \textbf {\bibinfo {volume} {22}},\ \bibinfo {pages} {737}
  (\bibinfo {year} {2004})}\BibitemShut {NoStop}%
\bibitem [{\citenamefont {Kouwenhoven}\ \emph {et~al.}(1990)\citenamefont
  {Kouwenhoven}, \citenamefont {{vAN WEEs}}, \citenamefont {{van der Vaart}},
  \citenamefont {Harmans}, \citenamefont {Timmering},\ and\ \citenamefont
  {Foxon}}]{kouwenhoven_selective_1990-2}%
  \BibitemOpen
  \bibfield  {author} {\bibinfo {author} {\bibfnamefont {L.~P.}\ \bibnamefont
  {Kouwenhoven}}, \bibinfo {author} {\bibfnamefont {B.~J.}\ \bibnamefont {{vAN
  WEEs}}}, \bibinfo {author} {\bibfnamefont {N.~C.}\ \bibnamefont {{van der
  Vaart}}}, \bibinfo {author} {\bibfnamefont {C.}~\bibnamefont {Harmans}},
  \bibinfo {author} {\bibfnamefont {C.~E.}\ \bibnamefont {Timmering}},\ and\
  \bibinfo {author} {\bibfnamefont {C.~T.}\ \bibnamefont {Foxon}},\ }\bibfield
  {title} {\bibinfo {title} {Selective population and detection of edge
  channels in the fractional quantum {{Hall}} regime},\ }\href@noop {}
  {\bibfield  {journal} {\bibinfo  {journal} {Physical review letters}\
  }\textbf {\bibinfo {volume} {64}},\ \bibinfo {pages} {685} (\bibinfo {year}
  {1990})}\BibitemShut {NoStop}%
\bibitem [{\citenamefont {Chang}\ and\ \citenamefont
  {Cunningham}(1992)}]{chang_transport_1992-1}%
  \BibitemOpen
  \bibfield  {author} {\bibinfo {author} {\bibfnamefont {A.~M.}\ \bibnamefont
  {Chang}}\ and\ \bibinfo {author} {\bibfnamefont {J.~E.}\ \bibnamefont
  {Cunningham}},\ }\bibfield  {title} {\bibinfo {title} {Transport evidence for
  phase separation into spatial regions of different fractional quantum
  {{Hall}} fluids near the boundary of a two-dimensional electron gas},\
  }\href@noop {} {\bibfield  {journal} {\bibinfo  {journal} {Physical review
  letters}\ }\textbf {\bibinfo {volume} {69}},\ \bibinfo {pages} {2114}
  (\bibinfo {year} {1992})}\BibitemShut {NoStop}%
\bibitem [{\citenamefont {Sabo}\ \emph {et~al.}(2017)\citenamefont {Sabo},
  \citenamefont {Gurman}, \citenamefont {Rosenblatt}, \citenamefont {Lafont},
  \citenamefont {Banitt}, \citenamefont {Park}, \citenamefont {Heiblum},
  \citenamefont {Gefen}, \citenamefont {Umansky},\ and\ \citenamefont
  {Mahalu}}]{sabo_edge_2017-3}%
  \BibitemOpen
  \bibfield  {author} {\bibinfo {author} {\bibfnamefont {R.}~\bibnamefont
  {Sabo}}, \bibinfo {author} {\bibfnamefont {I.}~\bibnamefont {Gurman}},
  \bibinfo {author} {\bibfnamefont {A.}~\bibnamefont {Rosenblatt}}, \bibinfo
  {author} {\bibfnamefont {F.}~\bibnamefont {Lafont}}, \bibinfo {author}
  {\bibfnamefont {D.}~\bibnamefont {Banitt}}, \bibinfo {author} {\bibfnamefont
  {J.}~\bibnamefont {Park}}, \bibinfo {author} {\bibfnamefont {M.}~\bibnamefont
  {Heiblum}}, \bibinfo {author} {\bibfnamefont {Y.}~\bibnamefont {Gefen}},
  \bibinfo {author} {\bibfnamefont {V.}~\bibnamefont {Umansky}},\ and\ \bibinfo
  {author} {\bibfnamefont {D.}~\bibnamefont {Mahalu}},\ }\bibfield  {title}
  {\bibinfo {title} {Edge reconstruction in fractional quantum {{Hall}}
  states},\ }\href {https://doi.org/10.1038/nphys4010} {\bibfield  {journal}
  {\bibinfo  {journal} {Nature Physics}\ }\textbf {\bibinfo {volume} {13}},\
  \bibinfo {pages} {491} (\bibinfo {year} {2017})}\BibitemShut {NoStop}%
\bibitem [{\citenamefont {Amet}\ \emph {et~al.}(2014)\citenamefont {Amet},
  \citenamefont {Williams}, \citenamefont {Watanabe}, \citenamefont
  {Taniguchi},\ and\ \citenamefont
  {{Goldhaber-Gordon}}}]{amet_selective_2014-1}%
  \BibitemOpen
  \bibfield  {author} {\bibinfo {author} {\bibfnamefont {F.}~\bibnamefont
  {Amet}}, \bibinfo {author} {\bibfnamefont {J.~R.}\ \bibnamefont {Williams}},
  \bibinfo {author} {\bibfnamefont {K.}~\bibnamefont {Watanabe}}, \bibinfo
  {author} {\bibfnamefont {T.}~\bibnamefont {Taniguchi}},\ and\ \bibinfo
  {author} {\bibfnamefont {D.}~\bibnamefont {{Goldhaber-Gordon}}},\ }\bibfield
  {title} {\bibinfo {title} {Selective {{Equilibration}} of
  {{Spin}}-{{Polarized Quantum Hall Edge States}} in {{Graphene}}},\ }\href
  {https://doi.org/10.1103/PhysRevLett.112.196601} {\bibfield  {journal}
  {\bibinfo  {journal} {Physical Review Letters}\ }\textbf {\bibinfo {volume}
  {112}},\ \bibinfo {pages} {196601} (\bibinfo {year} {2014})}\BibitemShut
  {NoStop}%
\bibitem [{\citenamefont {Wei}\ \emph {et~al.}(2017)\citenamefont {Wei},
  \citenamefont {{van der Sar}}, \citenamefont {{Sanchez-Yamagishi}},
  \citenamefont {Watanabe}, \citenamefont {Taniguchi}, \citenamefont
  {{Jarillo-Herrero}}, \citenamefont {Halperin},\ and\ \citenamefont
  {Yacoby}}]{wei_mach-zehnder_2017}%
  \BibitemOpen
  \bibfield  {author} {\bibinfo {author} {\bibfnamefont {D.~S.}\ \bibnamefont
  {Wei}}, \bibinfo {author} {\bibfnamefont {T.}~\bibnamefont {{van der Sar}}},
  \bibinfo {author} {\bibfnamefont {J.~D.}\ \bibnamefont
  {{Sanchez-Yamagishi}}}, \bibinfo {author} {\bibfnamefont {K.}~\bibnamefont
  {Watanabe}}, \bibinfo {author} {\bibfnamefont {T.}~\bibnamefont {Taniguchi}},
  \bibinfo {author} {\bibfnamefont {P.}~\bibnamefont {{Jarillo-Herrero}}},
  \bibinfo {author} {\bibfnamefont {B.~I.}\ \bibnamefont {Halperin}},\ and\
  \bibinfo {author} {\bibfnamefont {A.}~\bibnamefont {Yacoby}},\ }\bibfield
  {title} {\bibinfo {title} {Mach-{{Zehnder}} interferometry using spin- and
  valley-polarized quantum {{Hall}} edge states in graphene},\ }\href
  {https://doi.org/10.1126/sciadv.1700600} {\bibfield  {journal} {\bibinfo
  {journal} {Science Advances}\ }\textbf {\bibinfo {volume} {3}},\ \bibinfo
  {pages} {e1700600} (\bibinfo {year} {2017})}\BibitemShut {NoStop}%
\bibitem [{\citenamefont {Zimmermann}\ \emph {et~al.}(2017)\citenamefont
  {Zimmermann}, \citenamefont {Jordan}, \citenamefont {Gay}, \citenamefont
  {Watanabe}, \citenamefont {Taniguchi}, \citenamefont {Han}, \citenamefont
  {Bouchiat}, \citenamefont {Sellier},\ and\ \citenamefont
  {Sac{\'e}p{\'e}}}]{zimmermann_tunable_2017}%
  \BibitemOpen
  \bibfield  {author} {\bibinfo {author} {\bibfnamefont {K.}~\bibnamefont
  {Zimmermann}}, \bibinfo {author} {\bibfnamefont {A.}~\bibnamefont {Jordan}},
  \bibinfo {author} {\bibfnamefont {F.}~\bibnamefont {Gay}}, \bibinfo {author}
  {\bibfnamefont {K.}~\bibnamefont {Watanabe}}, \bibinfo {author}
  {\bibfnamefont {T.}~\bibnamefont {Taniguchi}}, \bibinfo {author}
  {\bibfnamefont {Z.}~\bibnamefont {Han}}, \bibinfo {author} {\bibfnamefont
  {V.}~\bibnamefont {Bouchiat}}, \bibinfo {author} {\bibfnamefont
  {H.}~\bibnamefont {Sellier}},\ and\ \bibinfo {author} {\bibfnamefont
  {B.}~\bibnamefont {Sac{\'e}p{\'e}}},\ }\bibfield  {title} {\bibinfo {title}
  {Tunable transmission of quantum {{Hall}} edge channels with full degeneracy
  lifting in split-gated graphene devices},\ }\href
  {https://doi.org/10.1038/ncomms14983} {\bibfield  {journal} {\bibinfo
  {journal} {Nature Communications}\ }\textbf {\bibinfo {volume} {8}},\
  \bibinfo {pages} {14983} (\bibinfo {year} {2017})}\BibitemShut {NoStop}%
\bibitem [{\citenamefont {Kumar}\ \emph {et~al.}(2018)\citenamefont {Kumar},
  \citenamefont {Srivastav},\ and\ \citenamefont
  {Das}}]{kumar_equilibration_2018}%
  \BibitemOpen
  \bibfield  {author} {\bibinfo {author} {\bibfnamefont {C.}~\bibnamefont
  {Kumar}}, \bibinfo {author} {\bibfnamefont {S.~K.}\ \bibnamefont
  {Srivastav}},\ and\ \bibinfo {author} {\bibfnamefont {A.}~\bibnamefont
  {Das}},\ }\bibfield  {title} {\bibinfo {title} {Equilibration of quantum
  {{Hall}} edges in symmetry-broken bilayer graphene},\ }\href
  {https://doi.org/10.1103/PhysRevB.98.155421} {\bibfield  {journal} {\bibinfo
  {journal} {Physical Review B}\ }\textbf {\bibinfo {volume} {98}},\ \bibinfo
  {pages} {155421} (\bibinfo {year} {2018})}\BibitemShut {NoStop}%
\bibitem [{\citenamefont {M{\"u}ller}\ \emph {et~al.}(1990)\citenamefont
  {M{\"u}ller}, \citenamefont {Weiss}, \citenamefont {Koch}, \citenamefont
  {{von Klitzing}}, \citenamefont {Nickel}, \citenamefont {Schlapp},\ and\
  \citenamefont {L{\"o}sch}}]{muller_edge_1990-2}%
  \BibitemOpen
  \bibfield  {author} {\bibinfo {author} {\bibfnamefont {G.}~\bibnamefont
  {M{\"u}ller}}, \bibinfo {author} {\bibfnamefont {D.}~\bibnamefont {Weiss}},
  \bibinfo {author} {\bibfnamefont {S.}~\bibnamefont {Koch}}, \bibinfo {author}
  {\bibfnamefont {K.}~\bibnamefont {{von Klitzing}}}, \bibinfo {author}
  {\bibfnamefont {H.}~\bibnamefont {Nickel}}, \bibinfo {author} {\bibfnamefont
  {W.}~\bibnamefont {Schlapp}},\ and\ \bibinfo {author} {\bibfnamefont
  {R.}~\bibnamefont {L{\"o}sch}},\ }\bibfield  {title} {\bibinfo {title} {Edge
  channels and the role of contacts in the quantum {{Hall}} regime},\
  }\href@noop {} {\bibfield  {journal} {\bibinfo  {journal} {Physical Review
  B}\ }\textbf {\bibinfo {volume} {42}},\ \bibinfo {pages} {7633} (\bibinfo
  {year} {1990})}\BibitemShut {NoStop}%
\bibitem [{\citenamefont {Faist}\ \emph {et~al.}(1991)\citenamefont {Faist},
  \citenamefont {Gu{\'e}ret},\ and\ \citenamefont
  {Meier}}]{faist_interior_1991}%
  \BibitemOpen
  \bibfield  {author} {\bibinfo {author} {\bibfnamefont {J.}~\bibnamefont
  {Faist}}, \bibinfo {author} {\bibfnamefont {P.}~\bibnamefont {Gu{\'e}ret}},\
  and\ \bibinfo {author} {\bibfnamefont {H.~P.}\ \bibnamefont {Meier}},\
  }\bibfield  {title} {\bibinfo {title} {Interior contacts for probing the
  equilibrium between magnetic edge channels in the quantum {{Hall}} effect},\
  }\href@noop {} {\bibfield  {journal} {\bibinfo  {journal} {Physical Review
  B}\ }\textbf {\bibinfo {volume} {43}},\ \bibinfo {pages} {9332} (\bibinfo
  {year} {1991})}\BibitemShut {NoStop}%
\bibitem [{\citenamefont {Woodside}\ \emph {et~al.}(2001)\citenamefont
  {Woodside}, \citenamefont {Vale}, \citenamefont {McEuen}, \citenamefont
  {Kadow}, \citenamefont {Maranowski},\ and\ \citenamefont
  {Gossard}}]{woodside_imaging_2001}%
  \BibitemOpen
  \bibfield  {author} {\bibinfo {author} {\bibfnamefont {M.~T.}\ \bibnamefont
  {Woodside}}, \bibinfo {author} {\bibfnamefont {C.}~\bibnamefont {Vale}},
  \bibinfo {author} {\bibfnamefont {P.~L.}\ \bibnamefont {McEuen}}, \bibinfo
  {author} {\bibfnamefont {C.}~\bibnamefont {Kadow}}, \bibinfo {author}
  {\bibfnamefont {K.~D.}\ \bibnamefont {Maranowski}},\ and\ \bibinfo {author}
  {\bibfnamefont {A.~C.}\ \bibnamefont {Gossard}},\ }\bibfield  {title}
  {\bibinfo {title} {Imaging interedge-state scattering centers in the quantum
  {{Hall}} regime},\ }\href {https://doi.org/10.1103/PhysRevB.64.041310}
  {\bibfield  {journal} {\bibinfo  {journal} {Physical Review B}\ }\textbf
  {\bibinfo {volume} {64}},\ \bibinfo {pages} {041310} (\bibinfo {year}
  {2001})}\BibitemShut {NoStop}%
\bibitem [{com({\natexlab{a}})}]{comm_topgates}%
  \BibitemOpen
  \href@noop {} {\bibinfo {title} {To be precise, in order to render the bulk
  density as uniform as possible, we energize the top gates with a small
  positive bias of {$\approx\SIrange{30}{50}{\milli\volt}$}. this counteracts
  the density modulation due to the presence alone of the gates on the surface
  of the device.}} ({\natexlab{a}})\BibitemShut {NoStop}%
\bibitem [{\citenamefont {Gerhardts}(2008)}]{gerhardts_effect_2008}%
  \BibitemOpen
  \bibfield  {author} {\bibinfo {author} {\bibfnamefont {R.~R.}\ \bibnamefont
  {Gerhardts}},\ }\bibfield  {title} {\bibinfo {title} {The effect of screening
  on current distribution and conductance quantisation in narrow quantum
  {{Hall}} systems},\ }\href {https://doi.org/10.1002/pssb.200743344}
  {\bibfield  {journal} {\bibinfo  {journal} {Physica Status Solidi (b)}\
  }\textbf {\bibinfo {volume} {245}},\ \bibinfo {pages} {378} (\bibinfo {year}
  {2008})}\BibitemShut {NoStop}%
\bibitem [{\citenamefont {Berl}\ \emph {et~al.}(2016)\citenamefont {Berl},
  \citenamefont {Tiemann}, \citenamefont {Dietsche}, \citenamefont {Karl},\
  and\ \citenamefont {Wegscheider}}]{berl_structured_2016}%
  \BibitemOpen
  \bibfield  {author} {\bibinfo {author} {\bibfnamefont {M.}~\bibnamefont
  {Berl}}, \bibinfo {author} {\bibfnamefont {L.}~\bibnamefont {Tiemann}},
  \bibinfo {author} {\bibfnamefont {W.}~\bibnamefont {Dietsche}}, \bibinfo
  {author} {\bibfnamefont {H.}~\bibnamefont {Karl}},\ and\ \bibinfo {author}
  {\bibfnamefont {W.}~\bibnamefont {Wegscheider}},\ }\bibfield  {title}
  {\bibinfo {title} {Structured back gates for high-mobility two-dimensional
  electron systems using oxygen ion implantation},\ }\href
  {https://doi.org/10.1063/1.4945090} {\bibfield  {journal} {\bibinfo
  {journal} {Applied Physics Letters}\ }\textbf {\bibinfo {volume} {108}},\
  \bibinfo {pages} {132102} (\bibinfo {year} {2016})}\BibitemShut {NoStop}%
\bibitem [{\citenamefont {Scharnetzky}\ \emph {et~al.}(2020)\citenamefont
  {Scharnetzky}, \citenamefont {Meyer}, \citenamefont {Berl}, \citenamefont
  {Reichl}, \citenamefont {Tiemann}, \citenamefont {Dietsche},\ and\
  \citenamefont {Wegscheider}}]{scharnetzky_novel_2020}%
  \BibitemOpen
  \bibfield  {author} {\bibinfo {author} {\bibfnamefont {J.}~\bibnamefont
  {Scharnetzky}}, \bibinfo {author} {\bibfnamefont {J.~M.}\ \bibnamefont
  {Meyer}}, \bibinfo {author} {\bibfnamefont {M.}~\bibnamefont {Berl}},
  \bibinfo {author} {\bibfnamefont {C.}~\bibnamefont {Reichl}}, \bibinfo
  {author} {\bibfnamefont {L.}~\bibnamefont {Tiemann}}, \bibinfo {author}
  {\bibfnamefont {W.}~\bibnamefont {Dietsche}},\ and\ \bibinfo {author}
  {\bibfnamefont {W.}~\bibnamefont {Wegscheider}},\ }\bibfield  {title}
  {\bibinfo {title} {A novel planar back-gate design to control the carrier
  concentrations in {{GaAs}}-based double quantum wells},\ }\href
  {https://doi.org/10.1088/1361-6641/ab9324} {\bibfield  {journal} {\bibinfo
  {journal} {Semiconductor Science and Technology}\ }\textbf {\bibinfo {volume}
  {35}},\ \bibinfo {pages} {085019} (\bibinfo {year} {2020})}\BibitemShut
  {NoStop}%
\bibitem [{\citenamefont {Nicol{\'i}}\ \emph {et~al.}(2019)\citenamefont
  {Nicol{\'i}}, \citenamefont {M{\"a}rki}, \citenamefont {Br{\"a}m},
  \citenamefont {R{\"o}{\"o}sli}, \citenamefont {Hennel}, \citenamefont
  {Hofmann}, \citenamefont {Reichl}, \citenamefont {Wegscheider}, \citenamefont
  {Ihn},\ and\ \citenamefont {Ensslin}}]{nicoli_quantum_2019}%
  \BibitemOpen
  \bibfield  {author} {\bibinfo {author} {\bibfnamefont {G.}~\bibnamefont
  {Nicol{\'i}}}, \bibinfo {author} {\bibfnamefont {P.}~\bibnamefont
  {M{\"a}rki}}, \bibinfo {author} {\bibfnamefont {B.~A.}\ \bibnamefont
  {Br{\"a}m}}, \bibinfo {author} {\bibfnamefont {M.~P.}\ \bibnamefont
  {R{\"o}{\"o}sli}}, \bibinfo {author} {\bibfnamefont {S.}~\bibnamefont
  {Hennel}}, \bibinfo {author} {\bibfnamefont {A.}~\bibnamefont {Hofmann}},
  \bibinfo {author} {\bibfnamefont {C.}~\bibnamefont {Reichl}}, \bibinfo
  {author} {\bibfnamefont {W.}~\bibnamefont {Wegscheider}}, \bibinfo {author}
  {\bibfnamefont {T.}~\bibnamefont {Ihn}},\ and\ \bibinfo {author}
  {\bibfnamefont {K.}~\bibnamefont {Ensslin}},\ }\bibfield  {title} {\bibinfo
  {title} {Quantum dot thermometry at ultra-low temperature in a dilution
  refrigerator with a {{4He}} immersion cell},\ }\href
  {https://doi.org/10.1063/1.5127830} {\bibfield  {journal} {\bibinfo
  {journal} {Review of Scientific Instruments}\ }\textbf {\bibinfo {volume}
  {90}},\ \bibinfo {pages} {113901} (\bibinfo {year} {2019})}\BibitemShut
  {NoStop}%
\bibitem [{com({\natexlab{b}})}]{comm_fridge}%
  \BibitemOpen
  \href@noop {} {\bibinfo {title} {Even though our setup in general could reach
  a lower base temperature, during this cooldown the setup suffered from a
  higher thermal load, likely due to a problem with the magnet switch heater.
  we decided to continue the investigation nevertheless as a lower temperature
  was not critical for this experiment.}} ({\natexlab{b}})\BibitemShut
  {NoStop}%
\bibitem [{\citenamefont {B{\"u}ttiker}(1986)}]{buttiker_four-terminal_1986-4}%
  \BibitemOpen
  \bibfield  {author} {\bibinfo {author} {\bibfnamefont {M.}~\bibnamefont
  {B{\"u}ttiker}},\ }\bibfield  {title} {\bibinfo {title} {Four-{{Terminal
  Phase}}-{{Coherent Conductance}}},\ }\href
  {https://doi.org/10.1103/PhysRevLett.57.1761} {\bibfield  {journal} {\bibinfo
   {journal} {Physical Review Letters}\ }\textbf {\bibinfo {volume} {57}},\
  \bibinfo {pages} {1761} (\bibinfo {year} {1986})}\BibitemShut {NoStop}%
\bibitem [{\citenamefont {Datta}(1995)}]{datta_electronic_1995}%
  \BibitemOpen
  \bibfield  {author} {\bibinfo {author} {\bibfnamefont {S.}~\bibnamefont
  {Datta}},\ }\href {https://doi.org/10.1017/CBO9780511805776} {\emph {\bibinfo
  {title} {Electronic {{Transport}} in {{Mesoscopic Systems}}}}},\ Cambridge
  {{Studies}} in {{Semiconductor Physics}} and {{Microelectronic Engineering}}\
  (\bibinfo  {publisher} {{Cambridge University Press}},\ \bibinfo {address}
  {{Cambridge}},\ \bibinfo {year} {1995})\BibitemShut {NoStop}%
\bibitem [{sup()}]{supmat}%
  \BibitemOpen
  \href@noop {} {\bibinfo {title} {See supplemental material.}}\BibitemShut
  {Stop}%
\bibitem [{com({\natexlab{c}})}]{comm_fluctuations}%
  \BibitemOpen
  \href@noop {} {\bibinfo {title} {Channels with {$\widetilde{\mu}_i^0 = 0$}
  end up sometimes having potential higher than channel 1. impurities may
  modulate the equilibration along the way such that channel 1 can first
  strongly equilibrate with another one with parallel spin and then partially
  decouple from it to scatter electrons with a spin-flip into a neighboring
  channel, resulting in channel 1 having lower electrochemical potential at the
  detector than some other channel with parallel spin.}}
  ({\natexlab{c}})\BibitemShut {NoStop}%
\bibitem [{\citenamefont {Chklovskii}\ \emph {et~al.}(1992)\citenamefont
  {Chklovskii}, \citenamefont {Shklovskii},\ and\ \citenamefont
  {Glazman}}]{chklovskii_electrostatics_1992-2}%
  \BibitemOpen
  \bibfield  {author} {\bibinfo {author} {\bibfnamefont {D.~B.}\ \bibnamefont
  {Chklovskii}}, \bibinfo {author} {\bibfnamefont {B.~I.}\ \bibnamefont
  {Shklovskii}},\ and\ \bibinfo {author} {\bibfnamefont {L.~I.}\ \bibnamefont
  {Glazman}},\ }\bibfield  {title} {\bibinfo {title} {Electrostatics of edge
  channels},\ }\href {https://doi.org/10.1103/PhysRevB.46.4026} {\bibfield
  {journal} {\bibinfo  {journal} {Physical Review B}\ }\textbf {\bibinfo
  {volume} {46}},\ \bibinfo {pages} {4026} (\bibinfo {year}
  {1992})}\BibitemShut {NoStop}%
\bibitem [{\citenamefont {Lier}\ and\ \citenamefont
  {Gerhardts}(1994)}]{lier_self-consistent_1994-1}%
  \BibitemOpen
  \bibfield  {author} {\bibinfo {author} {\bibfnamefont {K.}~\bibnamefont
  {Lier}}\ and\ \bibinfo {author} {\bibfnamefont {R.~R.}\ \bibnamefont
  {Gerhardts}},\ }\bibfield  {title} {\bibinfo {title} {Self-consistent
  calculations of edge channels in laterally confined two-dimensional electron
  systems},\ }\href {https://doi.org/10.1103/PhysRevB.50.7757} {\bibfield
  {journal} {\bibinfo  {journal} {Physical Review B}\ }\textbf {\bibinfo
  {volume} {50}},\ \bibinfo {pages} {7757} (\bibinfo {year}
  {1994})}\BibitemShut {NoStop}%
\bibitem [{\citenamefont {Kovrizhin}\ and\ \citenamefont
  {Chalker}(2011)}]{kovrizhin_equilibration_2011-1}%
  \BibitemOpen
  \bibfield  {author} {\bibinfo {author} {\bibfnamefont {D.~L.}\ \bibnamefont
  {Kovrizhin}}\ and\ \bibinfo {author} {\bibfnamefont {J.~T.}\ \bibnamefont
  {Chalker}},\ }\bibfield  {title} {\bibinfo {title} {Equilibration of integer
  quantum {{Hall}} edge states},\ }\href
  {https://doi.org/10.1103/PhysRevB.84.085105} {\bibfield  {journal} {\bibinfo
  {journal} {Physical Review B}\ }\textbf {\bibinfo {volume} {84}},\ \bibinfo
  {pages} {085105} (\bibinfo {year} {2011})}\BibitemShut {NoStop}%
\bibitem [{\citenamefont {M{\"u}ller}\ \emph
  {et~al.}(1992{\natexlab{c}})\citenamefont {M{\"u}ller}, \citenamefont
  {Weiss}, \citenamefont {{von Klitzing}}, \citenamefont {Ploog}, \citenamefont
  {Nickel}, \citenamefont {Schlapp},\ and\ \citenamefont
  {L{\"o}sch}}]{muller_confinement-potential_1992}%
  \BibitemOpen
  \bibfield  {author} {\bibinfo {author} {\bibfnamefont {G.}~\bibnamefont
  {M{\"u}ller}}, \bibinfo {author} {\bibfnamefont {D.}~\bibnamefont {Weiss}},
  \bibinfo {author} {\bibfnamefont {K.}~\bibnamefont {{von Klitzing}}},
  \bibinfo {author} {\bibfnamefont {K.}~\bibnamefont {Ploog}}, \bibinfo
  {author} {\bibfnamefont {H.}~\bibnamefont {Nickel}}, \bibinfo {author}
  {\bibfnamefont {W.}~\bibnamefont {Schlapp}},\ and\ \bibinfo {author}
  {\bibfnamefont {R.}~\bibnamefont {L{\"o}sch}},\ }\bibfield  {title} {\bibinfo
  {title} {Confinement-potential tuning: {{From}} nonlocal to local
  transport},\ }\href@noop {} {\bibfield  {journal} {\bibinfo  {journal}
  {Physical Review B}\ }\textbf {\bibinfo {volume} {46}},\ \bibinfo {pages}
  {4336} (\bibinfo {year} {1992}{\natexlab{c}})}\BibitemShut {NoStop}%
\bibitem [{\citenamefont {Hirai}\ \emph {et~al.}(1995)\citenamefont {Hirai},
  \citenamefont {Komiyama}, \citenamefont {Fukatsu}, \citenamefont {Osada},
  \citenamefont {Shiraki},\ and\ \citenamefont
  {Toyoshima}}]{hirai_dependence_1995-1}%
  \BibitemOpen
  \bibfield  {author} {\bibinfo {author} {\bibfnamefont {H.}~\bibnamefont
  {Hirai}}, \bibinfo {author} {\bibfnamefont {S.}~\bibnamefont {Komiyama}},
  \bibinfo {author} {\bibfnamefont {S.}~\bibnamefont {Fukatsu}}, \bibinfo
  {author} {\bibfnamefont {T.}~\bibnamefont {Osada}}, \bibinfo {author}
  {\bibfnamefont {Y.}~\bibnamefont {Shiraki}},\ and\ \bibinfo {author}
  {\bibfnamefont {H.}~\bibnamefont {Toyoshima}},\ }\bibfield  {title} {\bibinfo
  {title} {Dependence of inter-edge-channel scattering on temperature and
  magnetic field: {{Insight}} into the edge-confining potential},\ }\href
  {https://doi.org/10.1103/PhysRevB.52.11159} {\bibfield  {journal} {\bibinfo
  {journal} {Physical Review B}\ }\textbf {\bibinfo {volume} {52}},\ \bibinfo
  {pages} {11159} (\bibinfo {year} {1995})}\BibitemShut {NoStop}%
\bibitem [{\citenamefont {Roulleau}\ \emph {et~al.}(2008)\citenamefont
  {Roulleau}, \citenamefont {Portier}, \citenamefont {Roche}, \citenamefont
  {Cavanna}, \citenamefont {Faini}, \citenamefont {Gennser},\ and\
  \citenamefont {Mailly}}]{roulleau_direct_2008-2}%
  \BibitemOpen
  \bibfield  {author} {\bibinfo {author} {\bibfnamefont {P.}~\bibnamefont
  {Roulleau}}, \bibinfo {author} {\bibfnamefont {F.}~\bibnamefont {Portier}},
  \bibinfo {author} {\bibfnamefont {P.}~\bibnamefont {Roche}}, \bibinfo
  {author} {\bibfnamefont {A.}~\bibnamefont {Cavanna}}, \bibinfo {author}
  {\bibfnamefont {G.}~\bibnamefont {Faini}}, \bibinfo {author} {\bibfnamefont
  {U.}~\bibnamefont {Gennser}},\ and\ \bibinfo {author} {\bibfnamefont
  {D.}~\bibnamefont {Mailly}},\ }\bibfield  {title} {\bibinfo {title} {Direct
  {{Measurement}} of the {{Coherence Length}} of {{Edge States}} in the
  {{Integer Quantum Hall Regime}}},\ }\href
  {https://doi.org/10.1103/PhysRevLett.100.126802} {\bibfield  {journal}
  {\bibinfo  {journal} {Physical Review Letters}\ }\textbf {\bibinfo {volume}
  {100}},\ \bibinfo {pages} {126802} (\bibinfo {year} {2008})}\BibitemShut
  {NoStop}%
\bibitem [{\citenamefont {Duprez}\ \emph {et~al.}(2019)\citenamefont {Duprez},
  \citenamefont {Sivre}, \citenamefont {Anthore}, \citenamefont {Aassime},
  \citenamefont {Cavanna}, \citenamefont {Ouerghi}, \citenamefont {Gennser},\
  and\ \citenamefont {Pierre}}]{duprez_macroscopic_2019}%
  \BibitemOpen
  \bibfield  {author} {\bibinfo {author} {\bibfnamefont {H.}~\bibnamefont
  {Duprez}}, \bibinfo {author} {\bibfnamefont {E.}~\bibnamefont {Sivre}},
  \bibinfo {author} {\bibfnamefont {A.}~\bibnamefont {Anthore}}, \bibinfo
  {author} {\bibfnamefont {A.}~\bibnamefont {Aassime}}, \bibinfo {author}
  {\bibfnamefont {A.}~\bibnamefont {Cavanna}}, \bibinfo {author} {\bibfnamefont
  {A.}~\bibnamefont {Ouerghi}}, \bibinfo {author} {\bibfnamefont
  {U.}~\bibnamefont {Gennser}},\ and\ \bibinfo {author} {\bibfnamefont
  {F.}~\bibnamefont {Pierre}},\ }\bibfield  {title} {\bibinfo {title}
  {Macroscopic {{Electron Quantum Coherence}} in a {{Solid}}-{{State
  Circuit}}},\ }\href {https://doi.org/10.1103/PhysRevX.9.021030} {\bibfield
  {journal} {\bibinfo  {journal} {Physical Review X}\ }\textbf {\bibinfo
  {volume} {9}},\ \bibinfo {pages} {021030} (\bibinfo {year}
  {2019})}\BibitemShut {NoStop}%
\bibitem [{\citenamefont {Rosenblatt}\ \emph {et~al.}(2020)\citenamefont
  {Rosenblatt}, \citenamefont {Konyzheva}, \citenamefont {Lafont},
  \citenamefont {Schiller}, \citenamefont {Park}, \citenamefont {Snizhko},
  \citenamefont {Heiblum}, \citenamefont {Oreg},\ and\ \citenamefont
  {Umansky}}]{rosenblatt_energy_2020}%
  \BibitemOpen
  \bibfield  {author} {\bibinfo {author} {\bibfnamefont {A.}~\bibnamefont
  {Rosenblatt}}, \bibinfo {author} {\bibfnamefont {S.}~\bibnamefont
  {Konyzheva}}, \bibinfo {author} {\bibfnamefont {F.}~\bibnamefont {Lafont}},
  \bibinfo {author} {\bibfnamefont {N.}~\bibnamefont {Schiller}}, \bibinfo
  {author} {\bibfnamefont {J.}~\bibnamefont {Park}}, \bibinfo {author}
  {\bibfnamefont {K.}~\bibnamefont {Snizhko}}, \bibinfo {author} {\bibfnamefont
  {M.}~\bibnamefont {Heiblum}}, \bibinfo {author} {\bibfnamefont
  {Y.}~\bibnamefont {Oreg}},\ and\ \bibinfo {author} {\bibfnamefont
  {V.}~\bibnamefont {Umansky}},\ }\bibfield  {title} {\bibinfo {title} {Energy
  {{Relaxation}} in {{Edge Modes}} in the {{Quantum Hall Effect}}},\ }\href
  {https://doi.org/10.1103/PhysRevLett.125.256803} {\bibfield  {journal}
  {\bibinfo  {journal} {Physical Review Letters}\ }\textbf {\bibinfo {volume}
  {125}},\ \bibinfo {pages} {256803} (\bibinfo {year} {2020})}\BibitemShut
  {NoStop}%
\bibitem [{\citenamefont {R{\"o}{\"o}sli}\ \emph
  {et~al.}(2020{\natexlab{a}})\citenamefont {R{\"o}{\"o}sli}, \citenamefont
  {Brem}, \citenamefont {Kratochwil}, \citenamefont {Nicol{\'i}}, \citenamefont
  {Braem}, \citenamefont {Hennel}, \citenamefont {M{\"a}rki}, \citenamefont
  {Berl}, \citenamefont {Reichl}, \citenamefont {Wegscheider}, \citenamefont
  {Ensslin}, \citenamefont {Ihn},\ and\ \citenamefont
  {Rosenow}}]{roosli_observation_2020}%
  \BibitemOpen
  \bibfield  {author} {\bibinfo {author} {\bibfnamefont {M.~P.}\ \bibnamefont
  {R{\"o}{\"o}sli}}, \bibinfo {author} {\bibfnamefont {L.}~\bibnamefont
  {Brem}}, \bibinfo {author} {\bibfnamefont {B.}~\bibnamefont {Kratochwil}},
  \bibinfo {author} {\bibfnamefont {G.}~\bibnamefont {Nicol{\'i}}}, \bibinfo
  {author} {\bibfnamefont {B.~A.}\ \bibnamefont {Braem}}, \bibinfo {author}
  {\bibfnamefont {S.}~\bibnamefont {Hennel}}, \bibinfo {author} {\bibfnamefont
  {P.}~\bibnamefont {M{\"a}rki}}, \bibinfo {author} {\bibfnamefont
  {M.}~\bibnamefont {Berl}}, \bibinfo {author} {\bibfnamefont {C.}~\bibnamefont
  {Reichl}}, \bibinfo {author} {\bibfnamefont {W.}~\bibnamefont {Wegscheider}},
  \bibinfo {author} {\bibfnamefont {K.}~\bibnamefont {Ensslin}}, \bibinfo
  {author} {\bibfnamefont {T.}~\bibnamefont {Ihn}},\ and\ \bibinfo {author}
  {\bibfnamefont {B.}~\bibnamefont {Rosenow}},\ }\bibfield  {title} {\bibinfo
  {title} {Observation of quantum {{Hall}} interferometer phase jumps due to a
  change in the number of bulk quasiparticles},\ }\href
  {https://doi.org/10.1103/PhysRevB.101.125302} {\bibfield  {journal} {\bibinfo
   {journal} {Physical Review B}\ }\textbf {\bibinfo {volume} {101}},\ \bibinfo
  {pages} {125302} (\bibinfo {year} {2020}{\natexlab{a}})}\BibitemShut
  {NoStop}%
\bibitem [{\citenamefont {R{\"o}{\"o}sli}\ \emph
  {et~al.}(2020{\natexlab{b}})\citenamefont {R{\"o}{\"o}sli}, \citenamefont
  {Hug}, \citenamefont {Nicol{\'i}}, \citenamefont {M{\"a}rki}, \citenamefont
  {Reichl}, \citenamefont {Rosenow}, \citenamefont {Wegscheider}, \citenamefont
  {Ensslin},\ and\ \citenamefont {Ihn}}]{roosli_fractional_2020}%
  \BibitemOpen
  \bibfield  {author} {\bibinfo {author} {\bibfnamefont {M.~P.}\ \bibnamefont
  {R{\"o}{\"o}sli}}, \bibinfo {author} {\bibfnamefont {M.}~\bibnamefont {Hug}},
  \bibinfo {author} {\bibfnamefont {G.}~\bibnamefont {Nicol{\'i}}}, \bibinfo
  {author} {\bibfnamefont {P.}~\bibnamefont {M{\"a}rki}}, \bibinfo {author}
  {\bibfnamefont {C.}~\bibnamefont {Reichl}}, \bibinfo {author} {\bibfnamefont
  {B.}~\bibnamefont {Rosenow}}, \bibinfo {author} {\bibfnamefont
  {W.}~\bibnamefont {Wegscheider}}, \bibinfo {author} {\bibfnamefont
  {K.}~\bibnamefont {Ensslin}},\ and\ \bibinfo {author} {\bibfnamefont
  {T.}~\bibnamefont {Ihn}},\ }\bibfield  {title} {\bibinfo {title} {Fractional
  {{Coulomb}} blockade for quasiparticle tunneling between edge channels},\
  }\href@noop {} {\bibfield  {journal} {\bibinfo  {journal} {arXiv:2005.12723
  [cond-mat]}\ } (\bibinfo {year} {2020}{\natexlab{b}})},\ \Eprint
  {https://arxiv.org/abs/2005.12723} {arXiv:2005.12723 [cond-mat]} \BibitemShut
  {NoStop}%
\bibitem [{\citenamefont {Nakamura}\ \emph {et~al.}(2019)\citenamefont
  {Nakamura}, \citenamefont {Fallahi}, \citenamefont {Sahasrabudhe},
  \citenamefont {Rahman}, \citenamefont {Liang}, \citenamefont {Gardner},\ and\
  \citenamefont {Manfra}}]{nakamura_aharonovbohm_2019}%
  \BibitemOpen
  \bibfield  {author} {\bibinfo {author} {\bibfnamefont {J.}~\bibnamefont
  {Nakamura}}, \bibinfo {author} {\bibfnamefont {S.}~\bibnamefont {Fallahi}},
  \bibinfo {author} {\bibfnamefont {H.}~\bibnamefont {Sahasrabudhe}}, \bibinfo
  {author} {\bibfnamefont {R.}~\bibnamefont {Rahman}}, \bibinfo {author}
  {\bibfnamefont {S.}~\bibnamefont {Liang}}, \bibinfo {author} {\bibfnamefont
  {G.~C.}\ \bibnamefont {Gardner}},\ and\ \bibinfo {author} {\bibfnamefont
  {M.~J.}\ \bibnamefont {Manfra}},\ }\bibfield  {title} {\bibinfo {title}
  {Aharonov\textendash{{Bohm}} interference of fractional quantum {{Hall}} edge
  modes},\ }\href {https://doi.org/10.1038/s41567-019-0441-8} {\bibfield
  {journal} {\bibinfo  {journal} {Nature Physics}\ ,\ \bibinfo {pages} {1}}
  (\bibinfo {year} {2019})}\BibitemShut {NoStop}%
\bibitem [{\citenamefont {Bartolomei}\ \emph {et~al.}(2020)\citenamefont
  {Bartolomei}, \citenamefont {Kumar}, \citenamefont {Bisognin}, \citenamefont
  {Marguerite}, \citenamefont {Berroir}, \citenamefont {Bocquillon},
  \citenamefont {Pla{\c c}ais}, \citenamefont {Cavanna}, \citenamefont {Dong},
  \citenamefont {Gennser}, \citenamefont {Jin},\ and\ \citenamefont
  {F{\`e}ve}}]{bartolomei_fractional_2020}%
  \BibitemOpen
  \bibfield  {author} {\bibinfo {author} {\bibfnamefont {H.}~\bibnamefont
  {Bartolomei}}, \bibinfo {author} {\bibfnamefont {M.}~\bibnamefont {Kumar}},
  \bibinfo {author} {\bibfnamefont {R.}~\bibnamefont {Bisognin}}, \bibinfo
  {author} {\bibfnamefont {A.}~\bibnamefont {Marguerite}}, \bibinfo {author}
  {\bibfnamefont {J.-M.}\ \bibnamefont {Berroir}}, \bibinfo {author}
  {\bibfnamefont {E.}~\bibnamefont {Bocquillon}}, \bibinfo {author}
  {\bibfnamefont {B.}~\bibnamefont {Pla{\c c}ais}}, \bibinfo {author}
  {\bibfnamefont {A.}~\bibnamefont {Cavanna}}, \bibinfo {author} {\bibfnamefont
  {Q.}~\bibnamefont {Dong}}, \bibinfo {author} {\bibfnamefont {U.}~\bibnamefont
  {Gennser}}, \bibinfo {author} {\bibfnamefont {Y.}~\bibnamefont {Jin}},\ and\
  \bibinfo {author} {\bibfnamefont {G.}~\bibnamefont {F{\`e}ve}},\ }\bibfield
  {title} {\bibinfo {title} {Fractional statistics in anyon collisions},\
  }\href {https://doi.org/10.1126/science.aaz5601} {\bibfield  {journal}
  {\bibinfo  {journal} {Science}\ }\textbf {\bibinfo {volume} {368}},\ \bibinfo
  {pages} {173} (\bibinfo {year} {2020})}\BibitemShut {NoStop}%
\bibitem [{\citenamefont {Granger}\ \emph {et~al.}(2009)\citenamefont
  {Granger}, \citenamefont {Eisenstein},\ and\ \citenamefont
  {Reno}}]{granger_observation_2009-1}%
  \BibitemOpen
  \bibfield  {author} {\bibinfo {author} {\bibfnamefont {G.}~\bibnamefont
  {Granger}}, \bibinfo {author} {\bibfnamefont {J.~P.}\ \bibnamefont
  {Eisenstein}},\ and\ \bibinfo {author} {\bibfnamefont {J.~L.}\ \bibnamefont
  {Reno}},\ }\bibfield  {title} {\bibinfo {title} {Observation of {{Chiral Heat
  Transport}} in the {{Quantum Hall Regime}}},\ }\href
  {https://doi.org/10.1103/PhysRevLett.102.086803} {\bibfield  {journal}
  {\bibinfo  {journal} {Physical Review Letters}\ }\textbf {\bibinfo {volume}
  {102}},\ \bibinfo {pages} {086803} (\bibinfo {year} {2009})}\BibitemShut
  {NoStop}%
\bibitem [{\citenamefont {Venkatachalam}\ \emph {et~al.}(2012)\citenamefont
  {Venkatachalam}, \citenamefont {Hart}, \citenamefont {Pfeiffer},
  \citenamefont {West},\ and\ \citenamefont
  {Yacoby}}]{venkatachalam_local_2012-1}%
  \BibitemOpen
  \bibfield  {author} {\bibinfo {author} {\bibfnamefont {V.}~\bibnamefont
  {Venkatachalam}}, \bibinfo {author} {\bibfnamefont {S.}~\bibnamefont {Hart}},
  \bibinfo {author} {\bibfnamefont {L.}~\bibnamefont {Pfeiffer}}, \bibinfo
  {author} {\bibfnamefont {K.}~\bibnamefont {West}},\ and\ \bibinfo {author}
  {\bibfnamefont {A.}~\bibnamefont {Yacoby}},\ }\bibfield  {title} {\bibinfo
  {title} {Local thermometry of neutral modes on the quantum {{Hall}} edge},\
  }\href {https://doi.org/10.1038/nphys2384} {\bibfield  {journal} {\bibinfo
  {journal} {Nature Physics}\ }\textbf {\bibinfo {volume} {8}},\ \bibinfo
  {pages} {676} (\bibinfo {year} {2012})}\BibitemShut {NoStop}%
\bibitem [{\citenamefont {Banerjee}\ \emph {et~al.}(2017)\citenamefont
  {Banerjee}, \citenamefont {Heiblum}, \citenamefont {Rosenblatt},
  \citenamefont {Oreg}, \citenamefont {Feldman}, \citenamefont {Stern},\ and\
  \citenamefont {Umansky}}]{banerjee_observed_2017-2}%
  \BibitemOpen
  \bibfield  {author} {\bibinfo {author} {\bibfnamefont {M.}~\bibnamefont
  {Banerjee}}, \bibinfo {author} {\bibfnamefont {M.}~\bibnamefont {Heiblum}},
  \bibinfo {author} {\bibfnamefont {A.}~\bibnamefont {Rosenblatt}}, \bibinfo
  {author} {\bibfnamefont {Y.}~\bibnamefont {Oreg}}, \bibinfo {author}
  {\bibfnamefont {D.~E.}\ \bibnamefont {Feldman}}, \bibinfo {author}
  {\bibfnamefont {A.}~\bibnamefont {Stern}},\ and\ \bibinfo {author}
  {\bibfnamefont {V.}~\bibnamefont {Umansky}},\ }\bibfield  {title} {\bibinfo
  {title} {Observed quantization of anyonic heat flow},\ }\href
  {https://doi.org/10.1038/nature22052} {\bibfield  {journal} {\bibinfo
  {journal} {Nature}\ }\textbf {\bibinfo {volume} {545}},\ \bibinfo {pages}
  {75} (\bibinfo {year} {2017})}\BibitemShut {NoStop}%
\bibitem [{\citenamefont {Lafont}\ \emph {et~al.}(2019)\citenamefont {Lafont},
  \citenamefont {Rosenblatt}, \citenamefont {Heiblum},\ and\ \citenamefont
  {Umansky}}]{lafont_counter-propagating_2019}%
  \BibitemOpen
  \bibfield  {author} {\bibinfo {author} {\bibfnamefont {F.}~\bibnamefont
  {Lafont}}, \bibinfo {author} {\bibfnamefont {A.}~\bibnamefont {Rosenblatt}},
  \bibinfo {author} {\bibfnamefont {M.}~\bibnamefont {Heiblum}},\ and\ \bibinfo
  {author} {\bibfnamefont {V.}~\bibnamefont {Umansky}},\ }\bibfield  {title}
  {\bibinfo {title} {Counter-propagating charge transport in the quantum
  {{Hall}} effect regime},\ }\href {https://doi.org/10.1126/science.aar3766}
  {\bibfield  {journal} {\bibinfo  {journal} {Science}\ }\textbf {\bibinfo
  {volume} {363}},\ \bibinfo {pages} {54} (\bibinfo {year} {2019})}\BibitemShut
  {NoStop}%
\bibitem [{\citenamefont {Cohen}\ \emph {et~al.}(2019)\citenamefont {Cohen},
  \citenamefont {Ronen}, \citenamefont {Yang}, \citenamefont {Banitt},
  \citenamefont {Park}, \citenamefont {Heiblum}, \citenamefont {Mirlin},
  \citenamefont {Gefen},\ and\ \citenamefont
  {Umansky}}]{cohen_synthesizing_2019}%
  \BibitemOpen
  \bibfield  {author} {\bibinfo {author} {\bibfnamefont {Y.}~\bibnamefont
  {Cohen}}, \bibinfo {author} {\bibfnamefont {Y.}~\bibnamefont {Ronen}},
  \bibinfo {author} {\bibfnamefont {W.}~\bibnamefont {Yang}}, \bibinfo {author}
  {\bibfnamefont {D.}~\bibnamefont {Banitt}}, \bibinfo {author} {\bibfnamefont
  {J.}~\bibnamefont {Park}}, \bibinfo {author} {\bibfnamefont {M.}~\bibnamefont
  {Heiblum}}, \bibinfo {author} {\bibfnamefont {A.~D.}\ \bibnamefont {Mirlin}},
  \bibinfo {author} {\bibfnamefont {Y.}~\bibnamefont {Gefen}},\ and\ \bibinfo
  {author} {\bibfnamefont {V.}~\bibnamefont {Umansky}},\ }\bibfield  {title}
  {\bibinfo {title} {Synthesizing a {$\nu$} =2/3 fractional quantum {{Hall}}
  effect edge state from counter-propagating {$\nu$} =1 and {$\nu$} =1/3
  states},\ }\href {https://doi.org/10.1038/s41467-019-09920-5} {\bibfield
  {journal} {\bibinfo  {journal} {Nature Communications}\ }\textbf {\bibinfo
  {volume} {10}},\ \bibinfo {pages} {1920} (\bibinfo {year}
  {2019})}\BibitemShut {NoStop}%
\bibitem [{\citenamefont {Banerjee}\ \emph {et~al.}(2018)\citenamefont
  {Banerjee}, \citenamefont {Heiblum}, \citenamefont {Umansky}, \citenamefont
  {Feldman}, \citenamefont {Oreg},\ and\ \citenamefont
  {Stern}}]{banerjee_observation_2018-2}%
  \BibitemOpen
  \bibfield  {author} {\bibinfo {author} {\bibfnamefont {M.}~\bibnamefont
  {Banerjee}}, \bibinfo {author} {\bibfnamefont {M.}~\bibnamefont {Heiblum}},
  \bibinfo {author} {\bibfnamefont {V.}~\bibnamefont {Umansky}}, \bibinfo
  {author} {\bibfnamefont {D.~E.}\ \bibnamefont {Feldman}}, \bibinfo {author}
  {\bibfnamefont {Y.}~\bibnamefont {Oreg}},\ and\ \bibinfo {author}
  {\bibfnamefont {A.}~\bibnamefont {Stern}},\ }\bibfield  {title} {\bibinfo
  {title} {Observation of half-integer thermal {{Hall}} conductance},\ }\href
  {https://doi.org/10.1038/s41586-018-0184-1} {\bibfield  {journal} {\bibinfo
  {journal} {Nature}\ }\textbf {\bibinfo {volume} {559}},\ \bibinfo {pages}
  {205} (\bibinfo {year} {2018})}\BibitemShut {NoStop}%
\bibitem [{\citenamefont {Dutta}\ \emph {et~al.}(2021)\citenamefont {Dutta},
  \citenamefont {Yang}, \citenamefont {Melcer}, \citenamefont {Kundu},
  \citenamefont {Heiblum}, \citenamefont {Umansky}, \citenamefont {Oreg},
  \citenamefont {Stern},\ and\ \citenamefont {Mross}}]{dutta_novel_2021}%
  \BibitemOpen
  \bibfield  {author} {\bibinfo {author} {\bibfnamefont {B.}~\bibnamefont
  {Dutta}}, \bibinfo {author} {\bibfnamefont {W.}~\bibnamefont {Yang}},
  \bibinfo {author} {\bibfnamefont {R.~A.}\ \bibnamefont {Melcer}}, \bibinfo
  {author} {\bibfnamefont {H.~K.}\ \bibnamefont {Kundu}}, \bibinfo {author}
  {\bibfnamefont {M.}~\bibnamefont {Heiblum}}, \bibinfo {author} {\bibfnamefont
  {V.}~\bibnamefont {Umansky}}, \bibinfo {author} {\bibfnamefont
  {Y.}~\bibnamefont {Oreg}}, \bibinfo {author} {\bibfnamefont {A.}~\bibnamefont
  {Stern}},\ and\ \bibinfo {author} {\bibfnamefont {D.}~\bibnamefont {Mross}},\
  }\bibfield  {title} {\bibinfo {title} {Novel method distinguishing between
  competing topological orders},\ }\href@noop {} {\bibfield  {journal}
  {\bibinfo  {journal} {arXiv:2101.01419 [cond-mat, physics:quant-ph]}\ }
  (\bibinfo {year} {2021})},\ \Eprint {https://arxiv.org/abs/2101.01419}
  {arXiv:2101.01419 [cond-mat, physics:quant-ph]} \BibitemShut {NoStop}%
\end{thebibliography}

%

\end{document}